\documentclass[sigconf]{acmart}
\AtBeginDocument{%
  }

\setcopyright{acmlicensed}
\copyrightyear{2026}
\acmYear{2026}
\acmDOI{XXXXXXX.XXXXXXX}
\acmConference[CCS '26]{the 2026 ACM SIGSAC Conference on Computer and Communications Security}{November 15-19, 2026}{The Hague, The Netherlands}
\acmISBN{978-1-4503-XXXX-X/2026/06}

\usepackage{colortbl}
\usepackage{enumerate}
\usepackage{bm}
\usepackage{url}
\newcommand{\heading}[1]{{\vspace{3pt}\noindent{\textbf{#1}}}}

\begin{document}

\title{Larger Scale Offers Better Security in the Nakamoto-style Blockchain}




\author{Junjie Hu}
\affiliation{%
 \institution{Shanghai Jiao Tong University}
 \city{Shanghai}
 \country{China}}






\begin{abstract}

Security analyses of Nakamoto-style blockchains typically assume that adversarial nodes can coordinate instantaneously through private channels. This assumption substantially overestimates adversarial effectiveness in real-world deployments, where adversaries are geographically distributed and subject to non-negligible coordination and communication delays.

In this paper, we revisit the security of Nakamoto-style consensus under explicit adversarial coordination constraints. We first introduce a static delay model that treats adversarial internal communication latency as an independent parameter. Using queueing theoretic analysis, we derive the effective growth rate of adversarial private chains and obtain a corrected security threshold. The results show that adversarial coordination delay monotonically improves security and can exceed the classical 50\% bound when adversarial coordination is slower than honest block propagation. We then develop a dynamic delay corruption model that captures the joint efforts of network scale, communication delay, and probabilistic node corruption under bounded adversarial resources. Under this model, we prove that increasing network scale simultaneously amplifies adversarial coordination costs and dilutes adversarial power, leading to asymptotically improved security: the probability that adversarial power exceeds the security threshold converges to zero as the network grows. Finally, we show that private chain attacks remain optimal whthin the considered threat model. Coordination delay constrains adversarial effectiveness without introducing more powerful attacks. Our findings provide a theoretical foundation for optimizing consensus protocols and assessing the robustness of large-scale blockchains.

\end{abstract}




\maketitle

\section{Introduction}\label{section 1}
\subsection{Nakamoto-style Blockchain} Nakamoto-style blockchains, exemplified by Bitcoin \cite{BitcoinAPeer-to-PeerElectronicCashSystem}, Ethereum \cite{EthereumWhitepaper} and their derivatives \cite{PHANTOMGHOSTDAG, Litecoin, Bitcoincash, Bitcoin-NG, Fruitchain, Ouroboros}, are decentralized distributed ledgers that enable trustless value transfer through a consensus mechanism centered on the longest chain rule. At the core of such systems lies a network of validators, who compete to generate new blocks (via mechanisms such as Proof-of-Work (PoW), Proof-of-Stake (PoS), or similar ones) and extend the blockchain. Validators independently generate blocks containing transactions, with block creation governed by probabilistic rules (for instance, PoW requires solving cryptographic puzzles \cite{PricingviaProcessingorCombattingJunkMail}, while PoS depends on the proportion of stakes \cite{SecuringProof-of-StakeBlockchainProtocols}). Blocks are propagated through a peer-to-peer network, and synchronization delays among nodes (caused by network topology, bandwidth limitations, or network congestion) affect the speed at which the network reaches consensus. The security of Nakamoto-style blockchains hinges on the balance of resources between honest nodes and adversarial nodes. Classic models posit that if honest nodes control the majority of resources (e.g., hash power in PoW or stakes in PoS), the longest chain maintained by honest nodes will eventually outpace any adversarial chain, ensuring liveness and persistence \cite{BitcoinAPeer-to-PeerElectronicCashSystem, Majorityisnotenough, OntheSecurityandPerformanceofProofofWorkBlockchains}.

\subsection{Existing Security Models} Traditional security models for Nakamoto-style blockchains \cite{Majorityisnotenough, greedy-mine, PracticalSettlementBoundsforLongestChainConsensus, PracticalSettlementBoundsforProofofWorkBlockchains, TightConsistencyBoundsforBitcoin, LargerscaleNakamotostyleBlockchainsDontNecessarilyOfferBetterSecurity, abettermethodtoanalyzeblockchainconsistency, OnAnalysisoftheBitcoinandPrismBackboneProtocolsinSynchronousNetworks, Closelatency-securitytrade-offfortheNakamotoconsensus, AnalysisoftheBlockchainProtocolinAsynchronousNetworks, thebitcoinbackboneprotocol, TheSleepyModelofConsensus} focus on deriving conditions under which persistence and liveness are guaranteed. The foundational work by Nakamoto \cite{BitcoinAPeer-to-PeerElectronicCashSystem} established that security holds if honest nodes control more than 50\% of resources ($\beta < 0.5$), assuming instantaneous network synchronization. Subsequent studies relaxed this assumption by incorporating network delays (\(\Delta\)) among honest nodes. Models by Garay et al. \cite{thebitcoinbackboneprotocol} analyzed security under bounded delays, deriving conditions like \(\rho_{hon}(1 - 2(\Delta + 1)\rho_{hon}) > \rho_{adv}\) to guarantee consistency. Ren \cite{Bitcoin'slatencysecurityanalysismadesimple} modeled block generation as a Poisson process, capturing the tradeoff between honest/adversarial block rates and delays, with results such as \(\rho_{hon}e^{-2\rho_{hon}\Delta} > \rho_{adv}\) for security. Albrecht et al. \cite{LargerscaleNakamotostyleBlockchainsDontNecessarilyOfferBetterSecurity} highlighted that network scale introduces competing effects: larger networks increase honest nodes’ communication delays (\(\Delta(n) \in \Theta(\log n)\)) but dilute adversarial power by making corruption harder. However, these models retain critical simplifications that limit their realism, particularly regarding adversarial behavior. Specifically, existing models universally overlook the internal communication delay $\Delta_a$ among adversarial nodes caused by geographical distribution, network congestion, and topological heterogeneity. This omission leads to systematic deviations in theoretical analyses of security threshold $\beta^*$ from real-world scenarios.

\subsection{Limitations and Challenges} Traditional blockchain security analyses \cite{Majorityisnotenough,abettermethodtoanalyzeblockchainconsistency,Bitcoin'slatencysecurityanalysismadesimple,LongestChainConsensusUnderBandwidthConstraint,TightConsistencyBoundsforBitcoin,TheBitcoinBackboneProtocolwithChainsofVariableDifficulty,AnAnalysisofBlockchainConsistencyinAsynchronousNetworksDerivingaNeatBound,AnalysisoftheBlockchainProtocolinAsynchronousNetworks, securityhighratetransactionprocessinginbitcoin} implicitly assume that adversarial nodes can synchronize private chains instantaneously. This assumption is unrealistic in decentralized networks. When malicious miners operate across global regions, internal synchronization delays ($\Delta_a$) may reach hundreds of milliseconds or even seconds, significantly impairing the efficacy of private attacks \cite{HashSplit, Decreasingsecuritythresholdagainstdoublespendattackinnetworkswithslowsynchronization}. 

\heading{Reasons for non-zero delay among adversaries.} We take a real-world case involving botnets and cross-regional mining pools as an example. In the 2023 Monero botnet attack, as documented in the CISA AA23-086A report, the adversary utilized the XMRig malware to control 12,000 distributed mining machines across more than 30 countries worldwide for mining activities \cite{CISA}. It covers regions including North America, Europe, East Asia, and Southeast Asia to establish a mining botnet. This setup aimed to evade IP blocking in single regions and disperse the risk of hash rate detection. Measured data from CISA indicates that the internal block synchronization delay of this botnet ranged from 200 ms to 800 ms, while the average propagation delay of honest Monero nodes during the same period was 100 ms to 300 ms.

\begin{figure}[t]
  \centering
  \includegraphics[width=\linewidth]{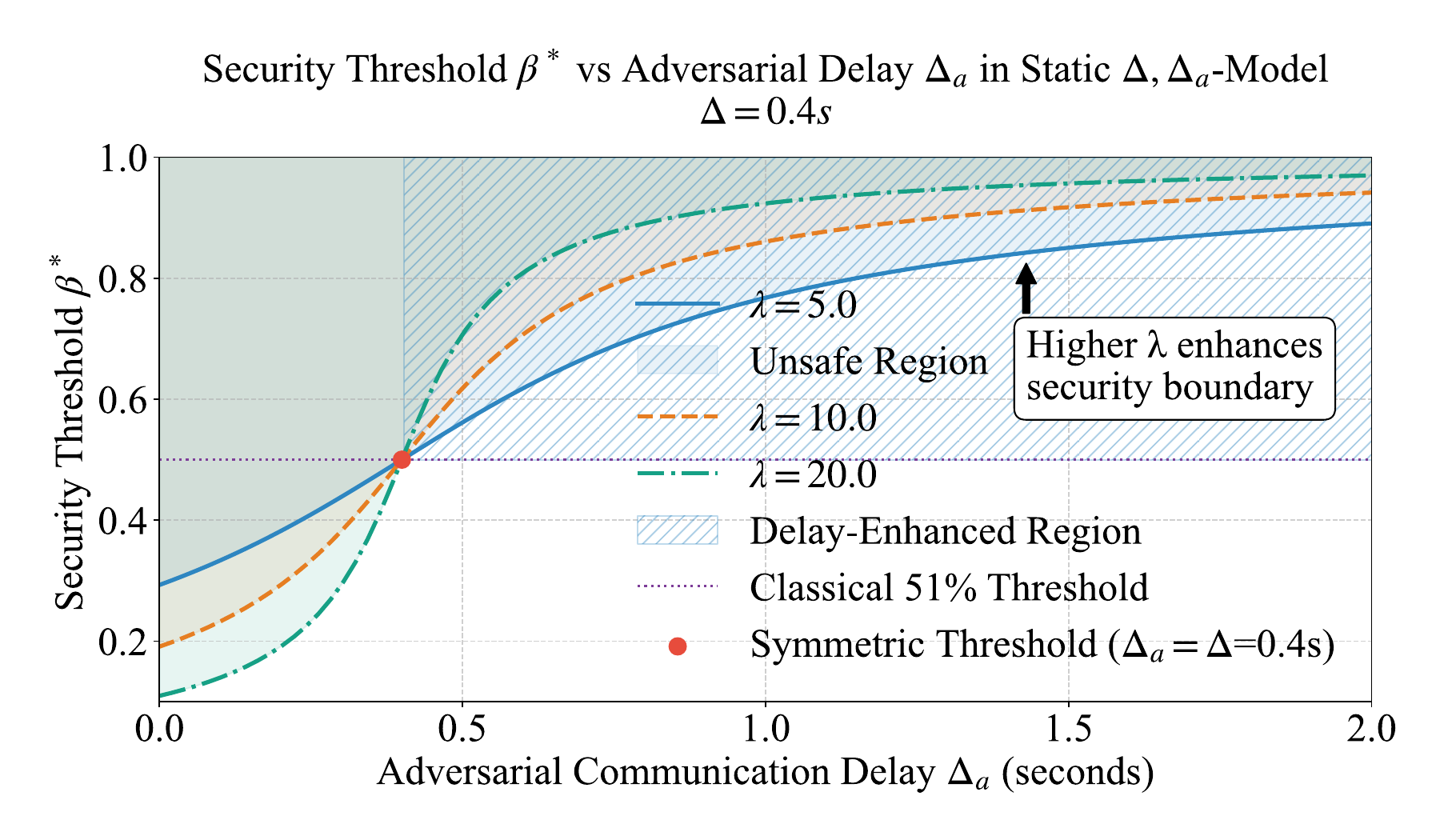}
  \caption{Security threshold comparison under adversarial communication delays in static $\Delta, \Delta_a$ model. This figure illustrates the relationship between the security threshold $\beta$ and the adversary communication delay $\Delta_a$ in a blockchain system. The horizontal axis represents $\Delta_a$ (0--2\,s), while the vertical axis denotes $\beta$ (0.1--1.0). Three curves with distinct styles correspond to block generation rates $\lambda = 5.0$, 10.0, and 20.0. All curves intersect at the symmetric threshold point $(0.4, 0.5)$ where $\Delta_a = \Delta = 0.4\,\mathrm{s}$. When $\Delta_a < 0.4\,\mathrm{s}$, $\beta$ remains below the classical 51\% threshold (purple dashed line). For $\Delta_a > 0.4\,\mathrm{s}$, $\beta$ increases with $\Delta_a$, showing stronger enhancement at higher $\lambda$. The diagonally shaded region ($\Delta_a > \Delta$) indicates latency amplification, while light regions above curves mark insecure configurations. Notably, $\Delta_a > \Delta$ enables superior security compared to traditional 51\% attacks, with higher $\lambda$ amplifying this enhancement through protocol-network interplay.}
  \label{Fig:1}
\end{figure}

\heading{Reasons for decentralized deployment.} In real-world scenarios, centralized deployment of adversaries entails non-negligible risks, while distributed deployment is an inevitable choice to balance efficiency and risk.  

For PoW blockchains, most countries worldwide impose regional hashrate restrictions on cryptocurrency mining. For instance, China implemented a ban on cryptocurrency mining in 2021, and the European Union (EU) introduced carbon emission caps for mining in 2024. If an adversary concentrates its hashrate in a single region, it is highly vulnerable to seizure by regulatory authorities. Additionally, centralized mining farms rely on a single set of power and network infrastructure. Once power outages or network disruptions occur (e.g., the 2021 Texas snowstorm that caused mining farm outages, during which Whinstone was completely shut down for nearly 8 days due to power failures), the adversary will instantly lose control of its hashrate \cite{Whinstone}. Consequently, leading mining pools generally adopt a deployment strategy combining core regions and distributed backups. For example, F2Pool has deployed backup hashrate across multiple regions globally, which directly results in a non-negligible adversarial internal delay (\(\Delta_a\)) \cite{F2Pool}.  

For PoS blockchains, Ethereum 2.0 strictly limits a single validator’s maximum stake to 32 ETHs through a maximum staking ratio constraint. This implies that an adversary must control a large number of independent validator nodes to obtain a higher proportion of stakes. If an adversary deploys validator nodes in a centralized manner, its IP ranges are likely to be blacklisted by honest nodes, preventing blocks from being propagated.

\begin{figure}[t]
  \centering
  \includegraphics[width=\linewidth]{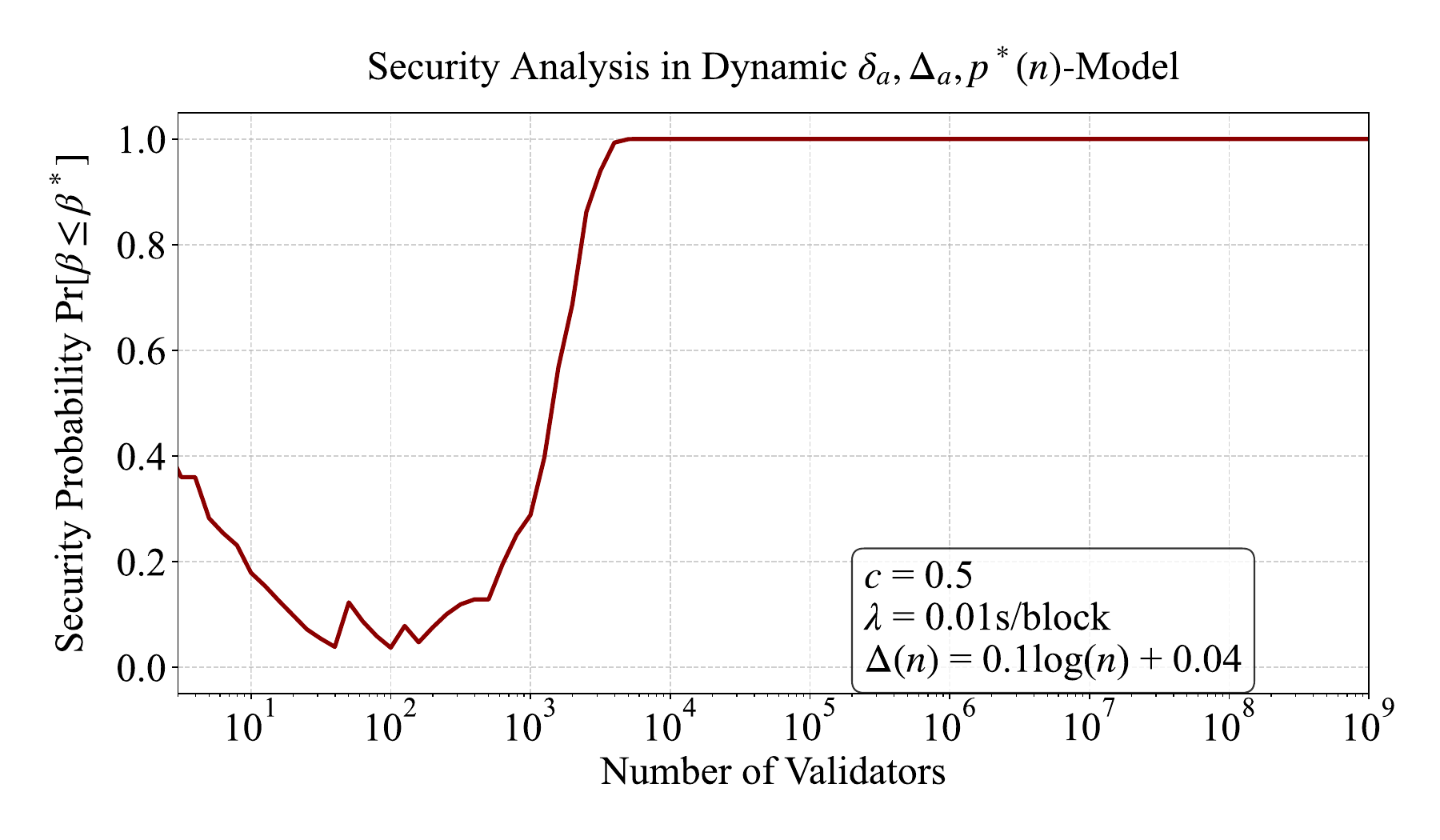}
  \caption{Security probability analysis in dynamic $\delta_a, \Delta_a, p^*(n)$ model. This figure compares security thresholds \(\beta^{*}\) under dynamic corruption rates, demonstrating how the number of validators (\(n\)) influences security probability. When the number of nodes is small, the network latency remains low, but the corruption ratio is relatively high. As the number of nodes increases, the network latency begins to grow logarithmically, which leads to a reduction in security probability. When the number of nodes continues to rise, the decrease in the corruption ratio outweighs the impact of increasing network latency, resulting in an improvement in security probability. Ultimately, when the number of nodes becomes very large, the corruption ratio diminishes to a negligible level despite persistent growth in network latency, thereby enhancing overall security probability. This non-monotonic trend (first decreasing then increasing) reflects the inherent trade-off relationship among network scale, latency, and corruption ratio in distributed systems.}
  \label{Fig:2}
\end{figure}

We summarize two limitations in the existing security models:

\begin{table}[H]
\centering
\begin{tabular}{|p{0.945\columnwidth}|}
\hline
\rowcolor[HTML]{EFEFEF}\textbf{L1: Overestimated Adversarial Coordination Efficiency.} \\ Existing models \cite{AnalysisoftheBlockchainProtocolinAsynchronousNetworks, abettermethodtoanalyzeblockchainconsistency} assume instantaneous adversarial synchronization ($\Delta_a = 0$), ignoring geographical and topological delays that hinder private chain propagation. \\
\rowcolor[HTML]{EFEFEF}\textbf{L2: Narrow Security Models.} \\ Prior analyses focus solely on honest node delays ($\Delta(n) \in \Theta(\log n)$) and adversarial power dilution, neglecting the interplay between network-scale-driven adversarial delays and honest block propagation dynamics. \\
\hline
\end{tabular}
\end{table}

These theoretical gaps leave two fundamental questions unresolved:

\begin{table}[H]
\centering
\begin{tabular}{|p{0.945\columnwidth}|}
\hline
\rowcolor[HTML]{EFEFEF}\textbf{Q1:} How does adversarial coordination delay ($\Delta_a$) independently affect the true security threshold $\beta^*$? \\
\rowcolor[HTML]{EFEFEF}\textbf{Q2:} Does increasing network scale ($n$) inherently enhance security, or does an optimal scale exist? \\
\hline
\end{tabular}
\end{table}

\subsection{Proposed Solutions} To address these gaps, we propose a dual-delay framework that explicitly models adversarial internal delays and their dynamic interaction with network scale:

    

\begin{table}[H]
\centering
\begin{tabular}{|p{0.945\columnwidth}|}
\hline
\rowcolor[HTML]{EFEFEF}\textbf{S1 to Q1:} 
\textbf{Static $\Delta, \Delta_a$ Model.} \\ We introduce adversarial communication delay \(\Delta_{a}\) to quantify synchronization inefficiencies. Using an M/D/1 queuing model, we derive the effective growth rate of adversarial chains. This approach aims to examine how \(\Delta_{a}\) influences the security threshold \(\beta^{*}\) and its relationship with the traditional boundary related to adversarial power proportion. \\
\hline
\rowcolor[HTML]{EFEFEF}\textbf{S2 to Q2:} 
\textbf{Dynamic $\delta_a, \Delta_a, p^*(n)$ Model.} \\ We incorporate probabilistic corruption, scale-dependent delays (where \(\Delta(n) \in \Theta(\log n)\)), and define the adversarial total delay window as \(\Delta_{total }=\Delta(n) \cdot e^{-k \beta}+c \cdot \log (1+\beta n)\). This model aims to analyze the interaction among network scale, communication delays, and adversarial corruption probability, exploring how these factors collectively influence the asymptotic behavior of adversarial power relative to the security threshold as the network scale grows. \\
\hline
\end{tabular}
\end{table}

\heading{Static $\Delta, \Delta_a$ Model.} The static $\Delta, \Delta_a$ model is an extension of Nakamoto's blockchain model, with its core innovation lying in the introduction of the communication delay \(\Delta_{a}\) among adversarial nodes to quantify the efficiency loss in the process of synchronizing private chains by adversarial nodes. This model derives the effective growth rate of the adversarial chain, \(\lambda_{eff} = \frac{\lambda_{a}}{1+\lambda_{a}\Delta_{a}}\), through the M/D/1 queuing model, a classic queuing theory model where M indicates that the process of block generation by adversarial nodes follows a Poisson distribution, D denotes that the service time for block synchronization is a fixed value \(\Delta_{a}\), and 1 represents a single server. This formula shows that the communication delay \(\Delta_{a}\) of adversarial nodes directly reduces the actual growth efficiency of their private chains: the larger \(\Delta_{a}\) is, the smaller the effective growth rate \(\lambda_{eff}\) of the adversarial chain becomes, meaning it is more difficult for adversarial nodes to surpass the main chain of honest nodes through their private chains. 

Based on this derivation, the model reveals the critical relationship between the communication delay \(\Delta_{a}\) of adversarial nodes and the security threshold \(\beta^{*}\) (the maximum proportion of adversarial node computing power that the system can tolerate): an increase in \(\Delta_{a}\) can significantly raise the security threshold \(\beta^{*}\). When the communication delay \(\Delta_{a}\) of adversarial nodes is greater than the communication delay \(\Delta\) of honest nodes, \(\beta^{*}\) may even exceed the boundary of the \textit{51\% attack} in traditional blockchain security models (i.e., \(\beta^{*} > 0.5\)). This implies that in scenarios where the synchronization efficiency of adversarial nodes is lower, the system may still remain secure even if adversarial nodes control more than 50\% of the computing power. 


\heading{Dynamic $\delta_a, \Delta_a, p^*(n)$ Model.} The Dynamic \(\delta_{a}\), \(\Delta_{a}\), \(p^{*}(n)\) model represents a dynamic extension to Nakamoto-style blockchain security analysis, with its core lying in the integration of probabilistic corruption, scale-dependent delays, and the adversarial total delay window, ultimately revealing the mechanism through which expanded network scale enhances security. The dynamic model characterizes the dynamic properties of real-world networks via probabilistic corruption, scale-dependent delays, and the adversarial total delay window. The probability of an adversarial node corrupting a single node is \(p^{*}(n)\), which decays as the network scale \(n\) increases. This is because a larger network requires adversaries to corrupt more nodes to gain sufficient computational power, and the probability of a single node being corrupted decreases due to power dilution. The communication delay among honest nodes, \(\Delta(n)\), grows logarithmically with the network scale \(n\) (\(\Delta(n) \in \Theta(\log n)\)). This is attributed to the increased geographical distribution and topological complexity as the number of network nodes rises, leading to prolonged information propagation delays. The total delay of adversarial nodes is decomposed into two components, given by the formula \(\Delta_{total }=\Delta(n) \cdot e^{-k \beta}+c \cdot \log (1+\beta n)\). The first component, \(\Delta(n) \cdot e^{-k \beta}\), represents the delay for adversarial nodes to receive blocks from honest nodes. It decreases exponentially as the adversarial computational power ratio \(\beta\) increases, as adversaries may acquire honest blocks more efficiently. The second component, \(c \cdot \log (1+\beta n)\), denotes the internal synchronization delay among adversarial nodes. It grows logarithmically with the network scale \(n\) and $\beta$, since a larger number of adversarial nodes or higher controlled computational power results in greater internal coordination costs.

A key conclusion of the dynamic model is that when the network scale \(n\) is sufficiently large, the combined effects of power dilution and delay suppression on adversarial nodes drive the system toward asymptotic security. On one hand, probabilistic corruption causes the expected value of the adversarial computational power ratio \(\beta\) to decay as \(n\) increases (\(\mathbb{E}[\beta]=p^{*}(n) \to 0\)), with its variance also decreasing, meaning \(\beta\) stabilizes at a low level. On the other hand, the total delay window \(\Delta_{total}\) of adversarial nodes expands with \(n\) (as both \(\Delta(n)\) and \(\log(1+\beta n)\) grow logarithmically), reducing the effective growth rate of the adversarial chain and making it harder to surpass the honest chain. As the network scale \(n \to \infty\), the model proves that \(\Pr[\beta \leq \beta^{*}] \to 1\), i.e., the probability that the adversarial computational power ratio \(\beta\) does not exceed the security threshold \(\beta^{*}\) approaches 1. This implies that the larger the network scale, the lower the likelihood of adversaries controlling sufficient computational power to launch an attack, and the more reliable the system's security. 

This conclusion rectifies the vague understanding of the relationship between network scale and security in traditional models, clarifying that decentralization (large-scale networks) achieves asymptotic improvement in security through the dual mechanisms of power dilution and delay suppression, ultimately demonstrating the superiority of large-scale Nakamoto-style blockchains.

\heading{Contributions.} Our core contributions are:
    
    

\begin{enumerate}[0]
    \item[$\bullet$] \textbf{Modeling adversarial coordination delay.} We introduce adversarial internal communication delay as a first-class parameter in Nakamoto-style security analysis, separating it from honest network delay and capturing coordination inefficiencies inherent in geographically distributed adversaries.
    
    \item[$\bullet$] \textbf{Static delay model and corrected security threshold.} We propose a static delay model incorporating adversarial coordination latency and derive the effective growth rate of adversarial private chains using queueing-theoretic analysis. This yields a closed-form approximation of the security threshold.
    
    \item[$\bullet$] \textbf{Dynamic delay–corruption model for scaling networks.} We develop a dynamic model that captures the joint effects of network scale, communication delay, and probabilistic node corruption under bounded adversarial resources. The model formalizes how scaling simultaneously increases adversarial coordination cost and dilutes adversarial power.

    \item[$\bullet$] \textbf{Asymptotic security characterization.} exceeding the security threshold converges to zero as the network size grows. This result provides a formal explanation for how decentralization can enhance security under realistic coordination constraints.

    \item[$\bullet$] \textbf{Optimality of private attacks under coordination delay.} We show that private attacks remain optimal adversarial strategies within the considered threat model, establishing that coordination delay constrains adversarial effectiveness without introducing more powerful attack classes.
\end{enumerate}

\section{Static \texorpdfstring{$\Delta, \Delta_a$} - Model}\label{section 2}

Traditional blockchain security models assume adversarial nodes can synchronize private chains instantaneously, leading to an overestimation of the security threshold. This assumption significantly deviates from real-world scenarios, particularly in globally distributed networks where such instantaneous coordination is infeasible. To systematically quantify the dynamic impact of adversarial internal communication delays on security, this section introduces the static $\Delta$, $\Delta_a$ model, which reconstructs the security analysis framework by incorporating maximum communication delay for block synchronization among adversarial nodes as an independent parameter.

\begin{figure}[t]
  \centering
  \includegraphics[width=\linewidth]{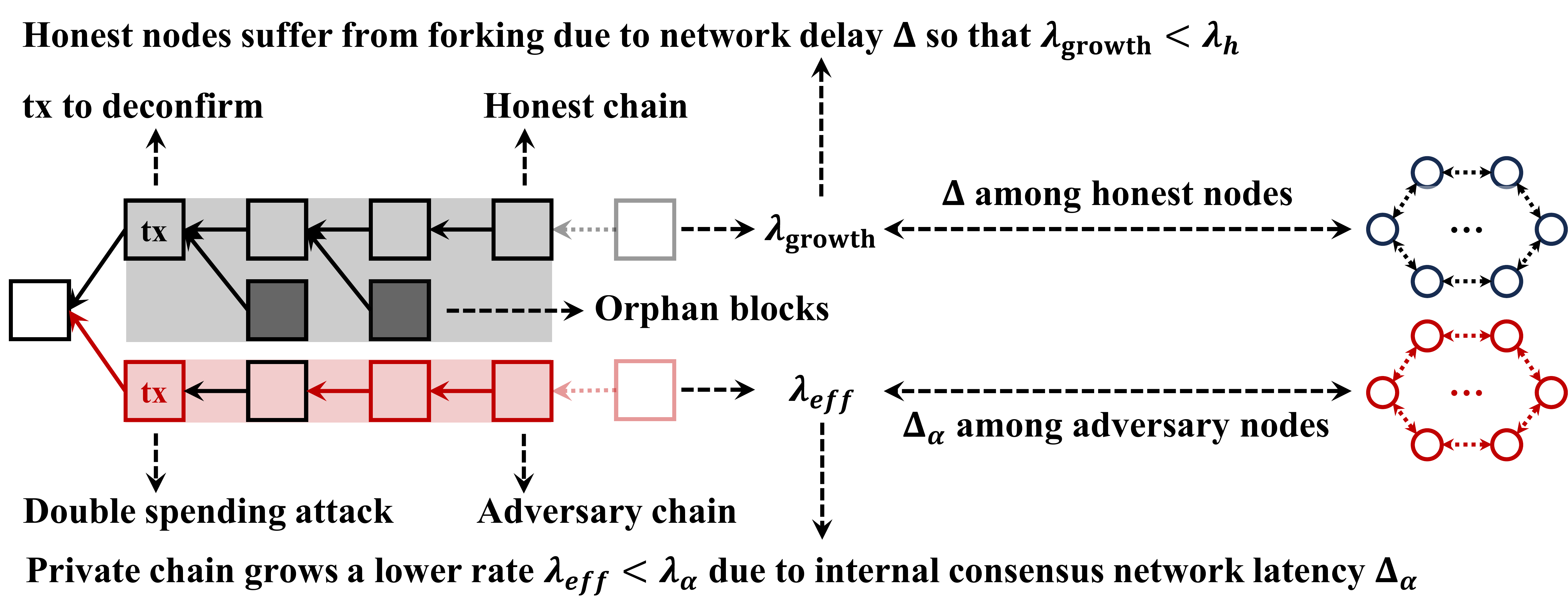}
  \caption{Static $\Delta, \Delta_{\alpha}$ model. This figure contrasts the growth of the honest chain (constrained by $\Delta$) and the adversarial private chain (constrained by $\Delta_{a}$), visually demonstrating how $\Delta_{a}$ suppresses the adversary’s chain growth rate.}
  \label{Fig:3}
\end{figure}

The model addresses two core questions: (1) how $\Delta_a$ independently regulates the security threshold, and (2) whether the system can withstand adversarial power exceeding the conventional 51\% attack boundary when $\Delta_a > \Delta$. To achieve this, we first define the model's notation and core assumptions in Section \ref{section 2.1}, then derive the effective chain growth rate using queuing theory and probabilistic analysis , and subsequently prove the optimality of private attack strategies in Section \ref{section 2.2}. Finally, we establish a quantitative relationship between $\Delta_a$ and $\beta$. This logical progression lays the theoretical foundation for the dynamic model presented in Section \ref{section 3}.

\subsection{Model and Assumptions}\label{section 2.1}

\subsubsection{Core Problem Restructuring}\label{section 2.1.1}
Traditional blockchain static security models are based on synchronous network assumptions, where adversarial nodes can achieve zero-delay collaboration. This idealized setting significantly deviates from the actual operating environment of decentralized networks. In real-world scenarios, adversarial nodes inevitably experience non-zero delays in internal block synchronization due to geographical distribution, network topology heterogeneity, and communication bandwidth limitations. However, existing research has not systematically quantified the dynamic impact of such delays on security thresholds. To address this theoretical gap, this study focuses on the coupling mechanism between internal adversarial communication delays and security thresholds, decomposing the core research problem into the following two progressive sub-problems:

\begin{table}[H]
\centering
\begin{tabular}{|p{0.945\columnwidth}|}
\hline
\rowcolor[HTML]{EFEFEF}\textbf{Q1: Static Delay Effect Analysis.} \\ How does the internal adversarial communication delay $\Delta_a$ affect the security threshold $\beta^*$? \\
\rowcolor[HTML]{EFEFEF}\textbf{Q2: Optimality of Private Attack Strategy.} \\ Is the private attack strategy still optimal in the static $\Delta, \Delta_a$ model framework? \\
\hline
\end{tabular}
\end{table}

The systematic resolution of these questions aims to reveal the quantitative relationship between adversarial collaboration efficiency and network delays, reconstruct a security evaluation framework aligned with actual network characteristics, and provide theoretical support for the robustness optimization of consensus protocols.

\subsubsection{System Model}\label{section 2.1.2}
We begin by defining the symbols employed in this section:

\begin{table}[H]
\centering
\begin{tabular}{cp{0.765\columnwidth}}
\toprule
\textbf{Symbol} & \textbf{Meaning} \\ \midrule
$n$       &    Total number of blockchain nodes, equal to the sum of validator nodes and zero-power nodes, $n = n_{\text{val}} + n_{\text{zp}}$.     \\
$\lambda_h$       &   Block generation rate of honest nodes (Poisson process).      \\
$\lambda_a$ & Block generation rate of adversarial nodes (Poisson process).\\
$\lambda$ & Total block generation rate, $\lambda = \lambda_h + \lambda_a$.\\
$\beta$ & Adversarial power proportion, $\beta = \frac{\lambda_a}{\lambda_a + \lambda_h}$.\\
$\Delta$ & Maximum communication delay among honest nodes.\\
$\Delta_a$ & Maximum communication delay for block synchronization among adversarial nodes.\\
$\lambda_{\text{eff}}$ & Effective growth rate of the adversarial chain.\\
$\lambda_{\text{growth}}$ & Effective growth rate of the honest chain.\\

\bottomrule
\end{tabular}
\end{table}

\heading{Node Sets.}
We partition blockchain nodes into validator nodes, which actively participate in generating new blocks and reaching consensus, and zero-power nodes, which only forward information and blocks, store a copy of the entire ledger, and help verify transactions but do not generate new blocks. We denote the set of all validators as $\mathcal{V}$, and the number of validators as $n_{\text{val}} = |\mathcal{V}|$. The number of zero-power nodes is denoted as $n_{\text{zp}}$, and the total number of nodes is $n = n_{\text{val}} + n_{\text{zp}}$. Without loss of generality, we assume that each validator possesses equal power \cite{Combiningghostandcasper}. Thus, the adversary's relative power can be characterized as the ratio of the number of validators it controls to the total number of validators in the system. It is worth noting that zero-power nodes also participate in block propagation. According to Ethereum PoS node statistics in November 2025, ignoring zero-power nodes would underestimate network latency by over 90\% \cite{EthereumPoS}. We further assume that all nodes are divided into honest nodes $\mathcal{V}_h$ and adversarial nodes $\mathcal{V}_a$, with the number of honest nodes $n_h = |\mathcal{V}_h|$ and adversarial nodes $n_a = |\mathcal{V}_a|$. We assume $\mathcal{V}$ can be represented as a sequence of key-value pairs, and use $i \in \mathcal{V}$ to denote the $i$-th validator in $\mathcal{V}$.

\heading{Block Generation Rate.}
Validators generate blocks at a rate of $\lambda$ blocks per second. If a new block appears at a certain time, it is generated by honest nodes following the longest chain rule at a rate of $\lambda_h$, or by adversarial nodes adopting malicious attack strategies at a rate of $\lambda_a$. Note that we assume $\lambda$ is a fixed system parameter, independent of the number of nodes $n$ or the number of validators $n_{\text{val}}$. In popular blockchain systems such as Bitcoin, $\lambda = \frac{1}{600}$, as Bitcoin periodically adjusts mining difficulty to maintain an average block generation time of approximately 10 minutes over the last 2016 blocks. Similarly, in Ethereum 1.0, $\lambda = \frac{1}{13}$, as Ethereum adjusts block difficulty to maintain an average block generation time of 13 seconds.

\heading{Static Communication Delay.}
Validators exchange information using a P2P network. Although in practice messages consist of blocks, transactions, and other metadata, our analysis focuses solely on block propagation. These blocks contain numerous transactions and require non-negligible time to transmit. Note that since adversarial nodes are mostly composed of mining pools, they use dedicated networks for message propagation, and their internal delays cannot be ignored. These delays differ significantly from the communication delays among honest nodes caused by geographical distribution and network congestion. To distinguish this difference and align with actual communication delays, we extend the traditional model, which only considers communication delays among honest nodes, to include internal communication delays among adversarial nodes. We use fixed values $\Delta$ and $\Delta_a$ to denote the maximum internal communication delay between honest nodes and adversarial nodes, respectively. Importantly, the growth of the adversarial chain is constrained by the synchronization of blocks generated before time $t - \Delta_a$. Similarly, the growth of the honest chain is constrained by blocks generated before time $t - \Delta$. Based on this analysis, we provide Definition \ref{Definition1}.

\begin{definition}[Adversarial Communication Delay $\bm\Delta_a$]\label{Definition1}
Building on the original model (M1-PoW/M1-PS/M1-Chia), we modify adversarial behavior as follows:
\begin{enumerate}[0]
    \item[$\bullet$] The communication delay among adversarial nodes is $\Delta_a$, meaning that private blocks published by adversarial nodes require $\Delta_a$ time to be received by other adversarial nodes.
    \item[$\bullet$] The communication delay among honest nodes remains $\Delta$, consistent with the original model.
\end{enumerate}
\end{definition}

\heading{Security Properties.}
Common security properties in the blockchain include liveness and persistence, which we formally define as follows:
\begin{enumerate}[0]
    \item[$\bullet$] \textbf{Liveness:} If a transaction is submitted to the network by an honest node at time $t$, and the network remains synchronized after $t$ (satisfying the partial synchrony assumption), there exists a finite time $T$ such that:
    \begin{enumerate}
        \item[(1)] \textbf{Availability:} For any time $t' > t + T$, the transaction is confirmed by all honest nodes.
    \end{enumerate}
    \item[$\bullet$] \textbf{Persistence:} If a transaction is confirmed by all honest nodes at time $t$ (i.e., included in a block on the local longest chain of honest nodes), then after time $t + \Delta$, the following holds:
    
\begin{enumerate}
        \item[(1)] \textbf{Consistency:} All honest nodes reach consensus on the confirmation status of the transaction.
        \item[(2)] \textbf{Irreversibility:} For any time $t' > t + \Delta$, no honest node considers the transaction revoked or modified.
    \end{enumerate}
\end{enumerate}

Note that liveness ensures that once a transaction is sufficiently confirmed by an honest node, it will be included in the blockchain copies of other honest nodes as the blockchain expands and propagates. This guarantees the consistency and stability of confirmed transactions across the entire blockchain network, preventing loss or tampering due to various network conditions (e.g., node failures, network delays, attacks), thereby maintaining the reliability and integrity of blockchain data.

Persistence ensures that in a blockchain system, any transaction received by honest nodes will not be indefinitely delayed or ignored but will be processed and added to the blockchain under certain conditions, becoming part of the network-wide recognized transaction records. This ensures the system's continuous processing and confirmation of transactions, preventing scenarios where transactions cannot be processed or remain in a pending state indefinitely, thereby maintaining the system's normal operation and functional effectiveness.

\subsubsection{Gap in Existing Static Security Models}\label{section 2.1.4}
Traditional static security models such as \cite{EverythingisaRaceandNakamotoAlwaysWins,UncleMaker,TightConsistencyBoundsforBitcoin}, when analyzing the security of blockchain protocols, generally rely on the following idealized assumption: adversarial nodes can achieve instantaneous internal collaboration. This setting significantly deviates from real-world decentralized network environments, leading to systematic deviations between theoretical analysis results and real-world security evaluations. Specifically, existing static security model research has critical theoretical flaws in the following two dimensions:

\begin{enumerate}[0]
    \item[$\bullet$] \textbf{Overestimation of Adversarial Collaboration Efficiency:} Existing models \cite{BitcoinAPeer-to-PeerElectronicCashSystem,Majorityisnotenough,EverythingisaRaceandNakamotoAlwaysWins} assume that adversarial nodes can synchronize blocks through delay-free private channels, ignoring the internal communication delays $\Delta_a$ caused by geographical distribution, topological heterogeneity, and bandwidth limitations. For example, when malicious miners are distributed across different regions, the synchronization delay of their private chains can reach hundreds of milliseconds, significantly reducing the effective growth rate of the adversarial chain. However, existing theoretical frameworks do not quantify the dynamic relationship between internal adversarial communication delays and security thresholds, leading to biased assessments of the actual attack capabilities of adversaries.
    
    \item[$\bullet$] \textbf{One-sidedness of Security Impact Mechanisms:} Although existing research \cite{BitcoinAPeer-to-PeerElectronicCashSystem,thebitcoinbackboneprotocol,abettermethodtoanalyzeblockchainconsistency} systematically analyzes the impact of honest node delays $\Delta$ on chain growth rates, it assumes that adversaries have perfect synchronization capabilities, failing to reveal the differential impact mechanisms of adversarial delays and honest delays. Specifically, when $\Delta_a$ and $\Delta$ differ significantly (e.g., adversarial mining pools using dedicated networks result in $\Delta_a < \Delta$), existing models cannot accurately capture the dynamic evolution of security thresholds $\beta^*$. Moreover, traditional models do not consider the potential impact of adversarial delays on the optimality of attack strategies, resulting in a lack of strategy universality in security condition derivations.
\end{enumerate}

The root cause of these theoretical flaws is that existing research does not model internal adversarial communication delays $\Delta_a$ as independent variables but conflates them with honest node delays $\Delta$ or ignores them entirely. This simplification not only overestimates the actual collaboration efficiency of adversaries but also causes the quantitative analysis of security thresholds $\beta^*$ to deviate from real-world network scenarios.

We incorporate $\Delta_a$ as an independent parameter into the static security model, constructing a quantitative framework for the effective growth rate of the adversarial chain $\lambda_{eff} = \frac{\lambda_a}{1 + \lambda_a \cdot \Delta_a}$, systematically revealing the dynamic regulatory mechanism of $\Delta_a$ on $\beta^*$.

\subsection{Security Analysis}\label{section 2.2}

Under the framework of the static $\Delta, \Delta_a$ model, this section investigates the dynamic regulatory mechanism of adversary internal communication delays on blockchain security through theoretical modeling and probabilistic analysis. We first establish a quantitative model for the effective growth rate of the adversary chain, derive the analytical expression of the security threshold $\beta^*$, and validate the theoretical completeness through the optimality proof of the private attack strategy.

We commence by employing Theorem \ref{Theorem1} to quantify the inhibitory effect of internal communication latency among adversaries on the rate of adversarial chain growth.

\begin{theorem}[Effective Adversarial Chain Growth Rate in the Static $\Delta, \Delta_a$ Model]\label{Theorem1}
In the static $\Delta, \Delta_a$ model, where:
\begin{itemize}
    \item Adversarial nodes generate blocks following a Poisson process with rate $\lambda_a$.
    \item Each adversarial block requires a fixed synchronization delay $\Delta_a$ to propagate among all adversarial nodes.
    \item The growth of the adversarial chain is constrained by blocks generated before time $t - \Delta_a$.
\end{itemize}
The effective growth rate of the adversarial chain, denoted by $\lambda_{eff}$, is given by:
\begin{equation}
    \lambda_{eff} = \frac{\lambda_a}{1 + \lambda_a \cdot \Delta_a}.
\end{equation}
\end{theorem}

Theorem \ref{Theorem1} formalizes the relationship between the internal adversarial communication delay $\Delta_a$ and the effective growth rate of the adversarial chain $\lambda_{eff}$. Specifically, it demonstrates that as $\Delta_a$ increases, the adversarial chain's growth rate decreases, reducing the adversary's ability to outpace the honest chain.

\begin{lemma}[System Utilization in M/D/1 Queue]\label{Lemma1}
For an M/D/1 queue modeling the adversarial synchronization process, where:
\begin{itemize}
    \item Block generation by adversarial nodes follows a Poisson process with rate $\lambda_a$.
    \item Each block requires a fixed synchronization delay $\Delta_a$ to propagate among all adversarial nodes.
\end{itemize}
The system utilization $\rho$ is defined as the product of the arrival rate $\lambda_a$ and the service time $\Delta_a$:
\begin{equation}
    \rho = \lambda_a \cdot \Delta_a
\end{equation}
where $\rho < 1$ ensures the stability of the queue.
\end{lemma}

The system utilization $\rho$ quantifies the efficiency of the adversarial synchronization process. When $\rho$ is close to 1, the queue is highly congested, leading to increased delays in block propagation and reduced effective growth rate of the adversarial chain. Conversely, when $\rho$ is small, the queue operates efficiently, allowing for faster synchronization and higher chain growth rates.

\begin{lemma}[Average Queue Length in M/D/1 Queue]\label{Lemma2}
For an M/D/1 queue modeling the adversarial synchronization process, where:
\begin{itemize}
    \item Block generation by adversarial nodes follows a Poisson process with rate $\lambda_a$.
    \item Each block requires a fixed synchronization delay $\Delta_a$ to propagate among all adversarial nodes.
    \item The average sojourn time $W$ represents the total time a block spends in the system (including both queuing and synchronization time).
\end{itemize}
The average queue length $L$, defined as the average number of blocks in the system (both in service and waiting), can be expressed using Little's Law as:
\begin{equation}
    L = \lambda_a \cdot W.
\end{equation}
\end{lemma}

The application of Little's Law in this context allows us to quantify the average number of blocks that are either being synchronized or waiting in the queue. When the arrival rate $\lambda_a$ is high or the synchronization delay $\Delta_a$ is significant, the average queue length $L$ increases, indicating a higher likelihood of delays in the adversarial chain growth. Conversely, a lower arrival rate or shorter synchronization delay reduces the queue length, leading to more efficient chain growth.

\begin{lemma}[Average Sojourn Time in M/D/1 Queue]\label{Lemma3}
For an M/D/1 queue modeling the adversarial synchronization process, where:
\begin{itemize}
    \item Block generation by adversarial nodes follows a Poisson process with rate $\lambda_a$.
    \item Each block requires a fixed synchronization delay $\Delta_a$ to propagate among all adversarial nodes.
    \item the system utilization $\rho = \lambda_a \cdot \Delta_a$.
\end{itemize}
The average sojourn time $W$, defined as the total time a block spends in the system (including both queuing and synchronization time), can be approximated as:
\begin{equation}
    W \approx \Delta_a + \frac{\rho \cdot \Delta_a}{2(1 - \rho)}.
\end{equation}
This approximation holds under steady-state conditions and assumes that the utilization $\rho$ is sufficiently small ($\rho \ll 1$), which is typical in practical blockchain deployments.
\end{lemma}

The average sojourn time $W$ quantifies the total delay experienced by a block in the adversarial synchronization process. When the system utilization $\rho$ is low, the queuing delay is minimal, and the sojourn time is dominated by the fixed synchronization delay $\Delta_a$. However, as $\rho$ increases, the queuing delay becomes significant, leading to a longer sojourn time.

\begin{theorem}[Security Condition and Threshold in the Static $\Delta, \Delta_a$ Model]\label{theorem2}
In the static $\Delta, \Delta_a$ model, where:
\begin{itemize}
    \item Honest nodes generate blocks at a rate $\lambda_h$ following a Poisson process.
    \item Adversarial nodes generate blocks at a rate $\lambda_a$ following a Poisson process.
    \item The total block generation rate $\lambda = \lambda_h + \lambda_a$.
    \item The maximum communication delay among honest nodes is $\Delta$.
    \item The maximum internal communication delay among adversarial nodes is $\Delta_a$.
\end{itemize}
The security threshold $\beta^*$, defined as the maximum fraction of adversarial power the system can tolerate while maintaining persistence and liveness, is given by:
\begin{equation}
    \beta^* \approx \frac{1}{2} + \frac{\lambda \cdot (\Delta_a - \Delta)}{4}.
\end{equation}
\end{theorem}

The modified security condition reveals that the adversarial communication delay $\Delta_a$ significantly influences the security threshold $\beta^*$. When $\Delta_a$ increases, the effective power of the adversary decreases, $\beta^*$ increases, and the system becomes more secure, capable of resisting adversaries with higher power. When $\Delta_a = \Delta$, $\beta^* = 0.5$, aligning with the symmetric case of honest nodes. When $\Delta_a \to 0$, the model reverts to the original result $\beta^* = \beta_{\text{pa}}$ (as $\lambda \cdot \Delta \to 0$).

Theorem \ref{theorem2} establishes a direct relationship between the internal adversarial communication delay $\Delta_a$ and the security threshold $\beta^*$. Specifically, it demonstrates that the larger the difference between $\Delta_a$ and $\Delta$, the higher the system's tolerance for adversarial power. By incorporating $\Delta_a$ as an independent parameter, the model provides a more accurate assessment of the system's resilience to attacks, especially in large-scale decentralized networks.

In the traditional static delay model, private attacks have been proven to be the optimal adversarial strategy. However, it remains uncertain whether this conclusion holds after introducing the adversarial internal communication delay $\Delta_a$. To address this question, we present Theorem \ref{Theorem3}, which formally proves that even in the presence of $\Delta_a$, private attacks remain the optimal strategy for the adversary, with their effective growth rate $\lambda_{eff}$ serving as the upper bound for all attack strategies. This conclusion not only provides strategic support for the security condition but also validates the applicability of private attack analysis in the context of the static $\Delta, \Delta_a$ model.

\begin{theorem}[Optimality of Private Attack in the Static $\Delta, \Delta_a$ Model]\label{Theorem3}
In the static $\Delta, \Delta_a$ model, where:
\begin{itemize}
    \item Honest nodes generate blocks following a Poisson process with rate $\lambda_h$.
    \item Adversarial nodes generate blocks following a Poisson process with rate $\lambda_a$.
    \item The maximum communication delay among honest nodes is $\Delta$.
    \item The maximum internal communication delay among adversarial nodes is $\Delta_a$.
\end{itemize}
The private attack strategy, where the adversary maintains a private chain and attempts to outpace the honest chain, is the optimal attack strategy.
\end{theorem}

Theorem \ref{Theorem3} formalizes the intuition that the private attack remains the most effective strategy for adversaries in the static $\Delta, \Delta_a$ model. This result demonstrates that, even when accounting for internal communication delays among adversarial nodes, the private attack provides the best possible chance for the adversary to compromise the blockchain's security. It also reveals that alternative attack strategies cannot outperform the private attack under realistic network conditions.

Next, we further explore the quantitative relationship between the security threshold $\beta^*$ and the adversarial delay $\Delta_a$. By applying the implicit function theorem, we prove that $\beta^*$ monotonically increases with $\Delta_a$, a property that validates the positive impact of adversarial internal communication delay on security.

\begin{lemma}[Monotonicity of Security Threshold $\beta^*$ with Respect to $\Delta_a$]\label{Lemma4}
In the static $\Delta, \Delta_a$ model, where:
\begin{itemize}
    \item Honest nodes generate blocks following a Poisson process with rate $\lambda_h$.
    \item Adversarial nodes generate blocks following a Poisson process with rate $\lambda_a$.
    \item The maximum communication delay among honest nodes is $\Delta$.
    \item The maximum internal communication delay among adversarial nodes is $\Delta_a$.
\end{itemize}
The security threshold $\beta^*$, defined as the maximum fraction of adversarial power the system can tolerate while maintaining persistence and liveness, is monotonically increasing with respect to $\Delta_a$.
\end{lemma}

Lemma \ref{Lemma4} formalizes the intuitive relationship between the internal adversarial communication delay $\Delta_a$ and the security threshold $\beta^*$. Specifically, it demonstrates that as $\Delta_a$ increases, the adversarial chain's growth rate is further throttled, reducing the adversary's ability to outpace the honest chain. This allows the system to tolerate a higher fraction of adversarial power, thereby increasing the security threshold $\beta^*$.

\begin{figure*}[t]
  \centering
  \includegraphics[width=0.9\linewidth]{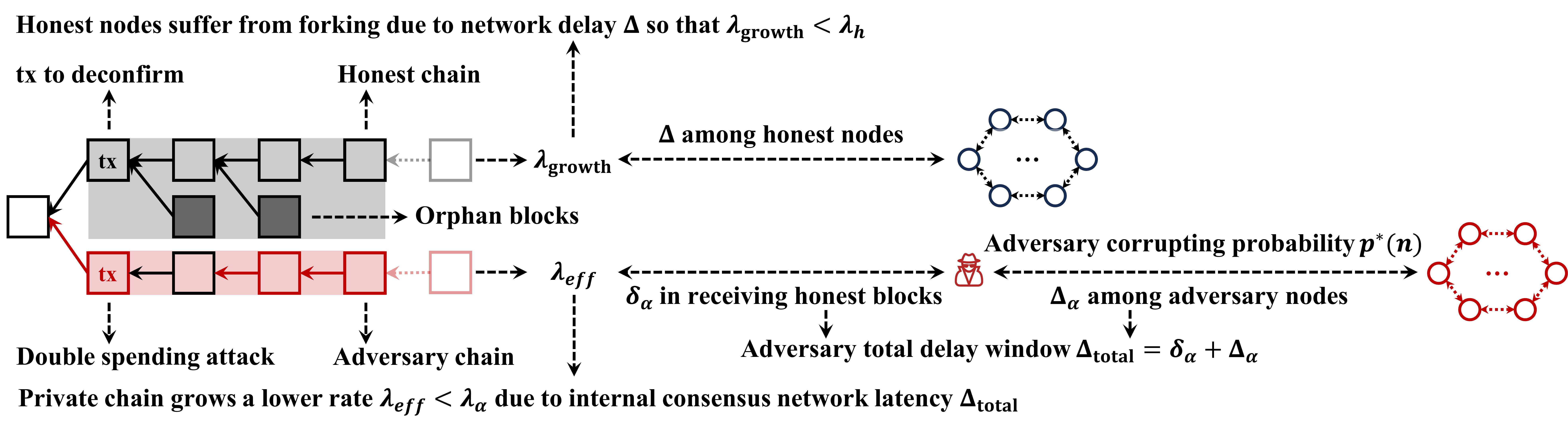}
  \caption{Dynamic $\delta_a, \Delta_a, p^*(n)$ model. This figure highlights the model’s key innovations: the decomposition of the adversary’s total delay into $\Delta_{\text{total}} = \delta_a + \Delta_a$, and the decay of the per-node corruption probability $p^*(n)$ with increasing network scale.}
  \label{Fig:4}
\end{figure*}

\section{Dynamic \texorpdfstring{$\delta_a, \Delta_a, p^*(n)$} - Model}\label{section 3}

The static model addresses the overestimation of adversarial coordination efficiency in traditional security threshold analyses by introducing the adversary's internal communication delay, revealing how latency differentials regulate security boundaries. However, its limitations lie in assuming fixed network size and adversary resource ratio, failing to capture the composite effects of dynamic network properties, such as node churn, network scaling, and the decay of adversarial corruption probability on security.

In real-world decentralized networks, scaling triggers two competing effects: (1) honest nodes' communication delay grows logarithmically, potentially reducing the honest chain's growth rate; (2) the adversary's corruption probability decays with network size, while its total delay window is dynamically coupled with network and adversary power ratio. Conventional dynamic models \cite{LargerscaleNakamotostyleBlockchainsDontNecessarilyOfferBetterSecurity} lack systematic analysis of these interactions, leading to biased asymptotic security assessments in large-scale networks.

To bridge this gap, this section proposes the dynamic $\delta_a, \Delta_a, p^*(n)$ model, establishing a unified analytical framework through three innovations:

\begin{enumerate}[0]
    \item[$\bullet$] \textbf{Dynamic Delay Decomposition}: The total adversarial delay is decomposed into (1) block reception delay, modeling how network scaling affects information acquisition efficiency, and (2) internal synchronization delay, quantifying coordination costs under resource constraints.
    
    \item[$\bullet$] \textbf{Corruption Probability Decay}: A node-wise independent corruption model formalizes the resource dilution effect, providing statistical convergence guarantees for asymptotic security.
    
    \item[$\bullet$] \textbf{Delay-Corruption Coupling Analysis}: By modeling the dynamic interplay between the adversary total delay window and the adversary power ratio, we demonstrate dual security enhancement pathways, latency suppression and corruption decay, ultimately proving the asymptotic property.
\end{enumerate}

The following subsections will formally define the model's assumptions and parameter system, followed by rigorous security analysis within this framework.

\subsection{Model and Assumptions}\label{section 3.1}

\subsubsection{Core Problem Reformulation}\label{section 3.1.1}

Traditional dynamic blockchain security models analyze the effects of network scaling under the assumption that adversary nodes can receive and synchronize honest blocks instantaneously, while treating the communication delays among honest nodes and the dilution of adversary power as independent processes. This theoretical simplification leads to two critical limitations: (1) insufficient modeling of the dynamic delays in adversary block reception and internal synchronization, resulting in an underestimation of the adversary's total delay window; and (2) failure to reveal the dynamic coupling mechanisms among network scale, communication delays, and adversary corruption probability, leaving the asymptotic security evaluation of large-scale networks without a unified framework. To address these theoretical gaps, this study decomposes the core research problem into the following three progressive objectives:

\begin{table}[H]
\centering
\begin{tabular}{|p{0.945\columnwidth}|}
\hline
\rowcolor[HTML]{EFEFEF}\textbf{Q1: Dynamic Delay Effect Analysis.} \\ How do dynamic delay models and the decay mechanism of adversary corruption probability affect the security threshold $\beta^*$? \\
\rowcolor[HTML]{EFEFEF}\textbf{Q2: Large-scale Asymptotic Security.} \\ Is the system asymptotically secure as the network size $n \rightarrow \infty$? \\
\hline
\end{tabular}
\end{table}

\begin{table}[H]
\centering
\begin{tabular}{|p{0.945\columnwidth}|}
\hline
\rowcolor[HTML]{EFEFEF}\textbf{Q3: Optimality of Private Attack Strategy.} \\ Does the private attack strategy remain optimal under the dynamic $\delta_a, \Delta_a, p^*(n)$ model framework? \\
\hline
\end{tabular}
\end{table}

The systematic resolution of these questions aims to construct a dynamic coupling model of network scale, communication delays, and adversary corruption probability, clarify the intrinsic mechanisms of decentralized security, and provide theoretical support for the robustness design of large-scale blockchain systems.

\subsubsection{System Model}\label{section 3.1.2}

We begin by defining the additional symbols employed in this section.

\begin{table}[H]
\centering
\begin{tabular}{cp{0.765\columnwidth}}
\toprule
\textbf{Symbol} & \textbf{Meaning} \\ \midrule
$\Delta(n)$       &    Maximum communication delay among honest nodes, $\Delta(n) \in \Theta(\log n)$.     \\
$\delta_a$ & Adversary's delay in receiving honest blocks, $\delta_a = \Delta(n) \ e^{-k\beta}$.\\
$\Delta_a$ & Adversary's internal synchronization delay, $\Delta_a = c \cdot \log(1+\beta \cdot n)$.\\
$\Delta_{\text{total}}$ & Adversary total delay window, $\Delta_{\text{total}} = \delta_a + \Delta_a$.\\
$\beta$ & Adversary power ratio, following a dynamic probability model, $\beta = \frac{\lambda_a}{\lambda_a + \lambda_h} = \frac{X}{n_{\text{val}}}$, where $X \sim \text{Binomial}(n_{\text{val}}, p^*)$.\\ 
$p^*(n)$ & Probability of adversary corrupting a single node, $p^*(n) = \frac{c}{\sqrt{n_{\text{val}}}}$.\\
\bottomrule
\end{tabular}
\end{table}

\heading{Dynamic Communication Delays.}
Traditional models (e.g., \cite{BitcoinAPeer-to-PeerElectronicCashSystem, Majorityisnotenough}) assume a fixed network scale, ignoring the dynamic impact of node count $n$ on the maximum communication delay $\Delta_n$ among honest nodes, the adversary's delay $\delta_a$ in receiving honest blocks, the adversary's internal synchronization delay $\Delta_a$, and the adversary's power $\beta$. However, in dynamic network scaling scenarios, the modeling of communication delays must comprehensively consider the dynamic impact of network scale expansion on the communication efficiency of both honest and adversary nodes. This model decomposes the delay parameters into three parts: honest node delay, adversary delay in receiving honest blocks, and adversary internal delay, defined as follows:

\begin{definition}[Honest Node Communication Delay $\Delta(n)$]\label{Definition2}
The maximum communication delay $\Delta(n)$ among honest nodes follows the diameter properties of Random Geometric Graphs (RGG) \cite{TheDiameterofSparseRandomGraphs}. Assuming $n$ nodes are uniformly distributed in a two-dimensional plane, the network diameter (communication delay) $\Delta(n)$ is:
\begin{equation}
    \Delta(n) = \Theta\left(\frac{\sqrt{\log n}}{\sqrt{\rho}}\right) = \Theta(\log n),
\end{equation}
where $\rho = \frac{n}{|S|}$ is the node density, and $|S|$ is the area of the region.
\end{definition}

Definition \ref{Definition2} indicates that as the number of nodes $n$ increases, $\Delta(n)$ grows logarithmically, reflecting the impact of scale expansion on propagation delay.

\begin{definition}[Adversary's Delay in Receiving Honest Blocks $\delta_a$]\label{Definition3}
Consider a blockchain with $n_{\text{total}} = n_{\text{val}} + n_{\text{zp}}$ nodes. The adversary's delay in receiving honest blocks follows a geometric distribution model:
\begin{equation}
    \delta_a = \Delta(n) \cdot e^{-k\beta},
\end{equation}
where:
\begin{itemize}
    \item $\Delta(n)$ is the maximum communication delay among honest nodes.
    \item $\beta = \frac{\lambda_a}{\lambda}$ is the adversary power ratio.
    \item $k > 0$ is a network topology constant (reflecting the efficiency of adversary power in reducing delay).
\end{itemize}
\end{definition}

$\delta_\alpha$ can be interpreted as the delay incurred by the adversary to monitor the length of the honest chain in order to optimize its attack strategy. This is a key parameter for maintaining the optimality of the private chain attack. Many studies have explicitly pointed out that an optimal private chain attack must monitor the honest chain so as to avoid over investment in a private chain that is doomed to fail \cite{Timestampmanipulation, HowtoBeatNakamotointheRace, Thehaltgame, Bitcoinundervolatileblockrewards, EfficientMDPanalysisforselfish-mininginblockchains, Optimalselfishminingstrategiesinbitcoin, Bitcoin'slatencysecurityanalysismadesimple,  Refinedbitcoinsecuritylatencyundernetworkdelay, SecurityLatencyandThroughputofProof-of-WorkNakamotoConsensus, Blockchainminingwithmultipleselfishminers, NovelBriberyMiningAttacks}. This monitoring process inevitably introduces a non‑zero delay $\delta_\alpha$, especially in large‑scale networks where honest block propagation is slow. $\delta_\alpha$ is crucial and cannot be ignored, because ignoring it would be equivalent to assuming an omniscient adversary that instantaneously knows the state of the honest chain. This scenario is not feasible in the real world.

\begin{definition}[Adversary's Internal Synchronization Delay $\Delta_a$]\label{Definition4}
Consider a blockchain with $n = n_{\text{val}} + n_{\text{zp}}$ nodes. The adversary's internal synchronization delay follows:
\begin{equation}
    \Delta_a = c \cdot \log(1+\beta \cdot n),
\end{equation}
where:
\begin{itemize}
    \item $n$ is the number of blockchain nodes.
    \item $\beta = \frac{\lambda_a}{\lambda}$ is the adversary power ratio.
    \item $c$ is a network topology constant.
\end{itemize}
\end{definition}

Definition \ref{Definition4} indicates that the communication delay among adversary nodes follows the same diameter properties of Random Geometric Graphs as honest nodes. Combining the above delays, the adversary total delay window $\Delta_{\text{total}}$ is:
\begin{equation}
    \Delta_{\text{total}} = \delta_a + \Delta_a = \Delta(n) \cdot e^{-k\beta} + c \cdot \log(1+\beta \cdot n).
\end{equation}

The adversary total delay window captures the combined delay in the adversary receiving honest blocks and internal synchronization, with its dynamic characteristics determined by the network scale $n$, adversary power $\beta$, and system parameters $k$ and $c$.

\heading{Adversary Corruption Model.}
Probabilistic corruption refers to adversaries randomly corrupting a subset of nodes through distributed infiltration methods, with typical scenarios including phishing attacks, vulnerability exploitation, and hashrate rental.
 
In 2023, the Pink Drainer phishing network launched large-scale phishing attacks targeting signature private keys and user wallets, with the objectives of disrupting the network and stealing user assets \cite{PinkDrainer, PinkDrainer1}. Notably, the probability of a single node being compromised in this attack was independent. In 2022, the Hezb crypto-mining malware actively exploited Remote Code Execution (RCE) vulnerabilities, notably in software like WSO2, to gain initial access to servers. This diverts the hijacked node's computational power for the attacker's illicit financial gain \cite{Mining}.

However, in large-scale networks, if an adversary intends to maintain control over a fixed proportion of nodes, it must invest resources that grow proportionally with the network scale. Consequently, for an adversary with a fixed amount of resources, the success rate of vulnerability-based attacks on individual nodes will systematically decrease as the network scale expands.
 
We provide a formal definition of the dynamic corruption model:

\begin{definition}[Adversarial Corruption Model]\label{Definition5}
The number of corrupted nodes $X$ follows a binomial distribution:
$$
X \sim \text{Binomial}(n, p)
$$
where:
\begin{itemize}
    \item $n$ is the total number of nodes in the blockchain network
    \item $p \in (0, 1)$ is the independent corruption probability for each node
\end{itemize}
The total corruption proportion is defined as:
\begin{equation}
    \beta = \frac{X}{n}.
\end{equation}
\end{definition}

Under the adversarial corruption model in Definition \ref{Definition5}, the corruption proportion $\beta$ satisfies:
\begin{equation}
    \mathbb{E}[\beta] = p \text{ and }\text{Var}(\beta) = \frac{p(1-p)}{n}.
\end{equation}
By the Law of Large Numbers, as $n \to \infty$, $ \beta \overset{}{\to} p. $

The adversary's probabilistic corruption model reflects the increasing difficulty of an attacker controlling validator nodes as the network scale grows in decentralized networks.

\subsubsection{Gaps in Existing Dynamic Security Models}\label{section 3.1.4}

Traditional dynamic blockchain security models (e.g., \cite{LargerscaleNakamotostyleBlockchainsDontNecessarilyOfferBetterSecurity,NakamotoConsensusunderBoundedProcessingCapacity,LongestChainConsensusUnderBandwidthConstraint}) exhibit critical theoretical deficiencies in two dimensions when analyzing the impact of network scale expansion on security: first, the inadequate modeling of the dynamic latency of adversary nodes; and second, the failure to reveal the coupling mechanism among network scale, communication latency, and adversary corruption probability. Specifically, the existing research frameworks suffer from the following systematic biases:

\begin{enumerate}[0]
    \item[$\bullet$] \textbf{Insufficient Modeling of Adversary Latency Dynamics.} Existing models generally assume that adversary nodes can instantly receive and synchronize honest blocks (i.e., ignoring the adversary's reception delay $\delta_a$) and that the adversary's internal communication delay $\Delta_a = 0$. This idealized setting significantly deviates from real-world network environments. On one hand, the adversary's process of receiving honest blocks is constrained by both the network scale $n$ and its own power ratio $\beta$, with the delay $\delta_a = \Delta(n) \cdot e^{-k\beta}$ decaying exponentially as $\beta$ increases. Existing studies \cite{LargerscaleNakamotostyleBlockchainsDontNecessarilyOfferBetterSecurity,NakamotoConsensusunderBoundedProcessingCapacity} fail to quantify the impact of this dynamic effect on the adversary's total delay window $\Delta_{\text{total}}$. On the other hand, the adversary's internal synchronization delay $\Delta_a$ grows logarithmically with the network scale $n$ and the adversary's power ratio $\beta$ ($\Delta_a \in \Theta(\log(\beta \cdot n))$), but \cite{LargerscaleNakamotostyleBlockchainsDontNecessarilyOfferBetterSecurity} conflates it with the honest delay $\Delta(n)$, leading to a systematic underestimation of the adversary's total delay window $\Delta_{\text{total}}$.

    \item[$\bullet$] \textbf{Insufficient Analysis of Dynamic Coupling Mechanisms.} Existing theoretical frameworks treat the decay of adversary power and the growth of latency for both honest and adversary nodes as independent processes, failing to construct a unified model to analyze their competing effects. On one hand, network scale expansion simultaneously induces negative effects ($\Delta(n)$ logarithmic growth reduces the growth rate of the honest chain) and positive effects (the decay of $\beta$ and the growth of $\Delta_{\text{total}}$ limit the adversary's capabilities). However, existing studies \cite{EverythingisaRaceandNakamotoAlwaysWins,TightConsistencyBoundsforBitcoin,LargerscaleNakamotostyleBlockchainsDontNecessarilyOfferBetterSecurity} do not reveal the dynamic balancing mechanism between these effects. On the other hand, the composite structure of the adversary's total delay window $\Delta_{\text{total}} = \Delta(n) \cdot e^{-k\beta} + c \cdot \log(1+\beta \cdot n)$ indicates a nonlinear interaction between the scale effect of $\Delta(n)$ and the corruption decay of $\beta$, while existing models \cite{AnalysisoftheBlockchainProtocolinAsynchronousNetworks} assume linear superposition, leading to asymptotic security analyses that deviate from real-world scenarios.
\end{enumerate}

By constructing a dynamic delay-corruption coupling model, We for the first time unify the interaction mechanisms of $\Delta(n)$, $\delta_a$, $\Delta_a$, and $\beta(n)$, systematically analyzing the asymptotic behavior of the adversary's total delay window and the competition between the decay rate of adversary power under network scale expansion. This theoretical breakthrough provides a rigorous mathematical foundation for the optimal scale design of decentralized networks and reveals the dual security enhancement paths of latency suppression and corruption decay.

\subsection{Security Analysis}\label{section 3.2}
Within the dynamic model framework, the expansion of network scale not only affects the communication latency of honest nodes but also profoundly influences system security through the decay of adversarial corruption probability and the dynamic coupling mechanism of delay windows. Building upon the monotonicity analysis between adversarial internal delay $\Delta_a$ and security threshold $\beta^{*}$ in the static model (Lemma \ref{Lemma4}), this section extends the research perspective to dynamic network environments, revealing the intrinsic relationships among network scale $n$, corruption probability $p^{*}(n)$, and delay parameters $\Delta(n)$, $\delta_{a}$, $\Delta_{a}$.

Specifically, the maximum communication delay $\Delta(n) \in \Theta(\log n)$ for honest nodes reflects the natural characteristic of network diameter growth with the number of nodes (Definition \ref{Definition2}). The adversary's delay in receiving honest blocks $\delta_{a} = \Delta(n)e^{-k\beta}$ (Definition \ref{Definition3}) indicates that as the adversarial power ratio $\beta$ increases, their efficiency in stealing honest blocks improves exponentially. This model captures the potential advantage adversaries gain through resource centralization to accelerate information acquisition. Concurrently, the adversary's internal synchronization delay $\Delta_{a} = c \cdot \log(1+\beta n)$ (Definition \ref{Definition4}) reveals the logarithmic decay of their collaboration efficiency with respect to network scale $n$ and their own resource ratio $\beta$. The total delay window $\Delta_{\text{total}} = \delta_{a} + \Delta_{a}$ combines bottlenecks in both information acquisition and internal synchronization phases, with its dynamic characteristics determined by network scale, adversarial power ratio, and system parameters.

The corruption probability model (Definition \ref{Definition5}) further reinforces decentralized security mechanisms: as the total number of validator nodes $n_{\text{val}}$ grows linearly with network expansion, the adversary must corrupt nodes through independent probability $p$. The variance of their expected power ratio $\operatorname{Var}(\beta) = \frac{p(1-p)}{n}$ decays as $n$ increases. By the Law of Large Numbers, when $n \rightarrow \infty$, $\beta \rightarrow p$, implying that adversaries struggle to significantly boost their power ratio through random fluctuations in large networks. Combined with the sublinear growth of delay parameters $\Delta_{\text{total}}$ (with $\Delta(n) \in \Theta(\log n)$ and $\Delta_{a} \in \Theta(\log n)$), the adversary's effective power is doubly suppressed—constrained both by the statistical convergence of corruption probability and the inflation of delay windows.

At this point, the dynamic model has fully characterized the tripartite coupling effects of network scale, delay mechanisms, and corruption decay. Theorem \ref{Theorem4} below will rigorously prove that as $n \rightarrow \infty$, the adversarial power ratio $\beta$ converges below the security threshold $\beta^{*}$ with probability approaching 1, thereby providing theoretical guarantees for the asymptotic security of large-scale blockchain systems.

\begin{theorem}[Security Condition and Threshold in the Dynamic $\delta_a, \Delta_a, p^*$ Model]\label{Theorem4}
In the dynamic $\delta_a, \Delta_a, p^*(n)$ model, where:
\begin{itemize}
    \item $p^*$ is the expected corruption proportion.
    \item $\beta^* = \frac{\lambda_{\text{eff}}}{\lambda_{\text{eff}} + \lambda_{\text{growth}}}$ is the security threshold, with $\lambda_{\text{eff}} = \frac{\lambda_a}{1 + \lambda_a \Delta_{\text{total}}}$.
    \item $\Delta_{\text{total}} = \Delta(n) \cdot e^{-k p^*} + c \cdot \log(1 + p^* n_{\text{val}})$ models the total adversarial delay window, where $\Delta(n) \in \Theta(\log n)$.
\end{itemize}
The security threshold $\beta^*$ is given by:
\begin{equation}
    \beta^* = \Theta\left(\frac{1}{\log n}\right).
\end{equation}
\end{theorem}

Theorem \ref{Theorem4} establishes the deterministic relationship of the security threshold $\beta^* = \Theta\left(\frac{1}{\sqrt{\log n}}\right)$ under the dynamic model. However, in practical systems, the adversarial power ratio $\beta$ is a random variable determined by the probabilistic corruption process. Specifically, $\beta = \frac{X}{n}$, where $X \sim \text{Binomial}(n, p^*(n))$ and $p^*(n) = \frac{c}{\sqrt{n}}$ denotes the node corruption probability. This probabilistic nature causes the actual distribution of $\beta$ to deviate from the theoretical threshold $\beta^*$, thus necessitating a further analysis of its probabilistic convergence.

Although Theorem \ref{Theorem4} provides the theoretical expression for the security threshold, the stochasticity of $\beta$ in the dynamic model arises from the probabilistic corruption of nodes. Specifically, the corruption probability $p^*(n)$ of each node decays with increasing network size $n$ ($p^*(n) = \frac{c}{\sqrt{n}}$), leading the expected value $\mathbb{E}[\beta] = p^*(n)$ to asymptotically approach zero. However, since the variance of the binomial distribution $\mathrm{Var}(\beta) = \frac{p^*(n)(1 - p^*(n))}{n}$ also decreases with $n$, the actual values of $\beta$ will concentrate tightly around the expected value. To quantify the probability that $\beta$ exceeds the security threshold $\beta^*$, we further analyze its distributional properties using probability theory, which forms the core of Theorem \ref{Theorem5}.

\begin{theorem}[Security Probability Expression in the Dynamic $\delta_a, \Delta_a, p^*$ Model]\label{Theorem5}
In the dynamic $\delta_a,\Delta_a,p^*(n)$ model, where:
\begin{itemize}
    \item $\Delta_{\text{total}} = \Delta(n) \cdot e^{-k p^*} + c \cdot \log(1 + p^* n_{\text{val}})$ models the total adversarial delay window, where $\Delta(n) \in \Theta(\log n)$.
    \item $\Phi(\cdot)$ is the standard normal cumulative distribution function (CDF).
    \item $p^*(n) = \frac{c}{\sqrt{n}}$ denotes the corruption probability;
    \item $\beta^* = \Theta\left(\frac{1}{\log n}\right)$ represents the security threshold.
\end{itemize}
The probability of adversarial power $\beta$ exceeding the security threshold $\beta^*$ is given by:
\begin{equation}
    \Pr[\beta > \beta^*] = 1 - \Phi\left(\sqrt{n} \cdot \frac{\beta^* - p^*(n)}{\sqrt{p^*(n)(1 - p^*(n))}}\right).
\end{equation}
\end{theorem}

Theorem \ref{Theorem5} provides a precise probabilistic characterization of adversarial success, yet its interpretation necessitates a deeper analysis of the standardized threshold $ Z_n $. This threshold quantifies the normalized separation between the security boundary $ \beta^* $ and the expected adversarial power $ p^*(n) $, scaled by the inherent variability in corruption outcomes. While both $ \beta^* $ and $ p^*(n) $ exhibit decay with increasing network size $ n $, their underlying decay mechanisms differ fundamentally. The security threshold $ \beta^* $ decays logarithmically due to delay mechanisms that amplify adversarial coordination costs. The corruption probability $ p^*(n)$ decays polynomially as node corruption becomes increasingly diluted with network scale.

This structural mismatch ensures that $ \beta^* $ remains orders of magnitude larger than $ p^*(n) $ for sufficiently large $ n $. When combined with the $ \sqrt{n} $-scaling factor in $ Z_n $, this disparity drives a superlinear growth in $ Z_n $.
As $ Z_n $ diverges, the Gaussian tail probability $ 1 - \Phi(Z_n) $ decays faster than any polynomial function, thereby guaranteeing $ \Pr[\beta > \beta^*] \to 0 $. This result mathematically formalizes the dual security mechanisms in large-scale networks: Communication delays systematically elevate the security threshold $ \beta^* $ and scale-dependent corruption probability $ p^*(n) $ exponentially suppresses adversarial power concentration.

\begin{theorem}[Security Probability Convergence at Scale in the Dynamic $\delta_a, \Delta_a, p^*$ Model]\label{Theorem6}
In the dynamic $\delta_a,\Delta_a,p^*(n)$ model, where:
\begin{itemize}
    \item $\Delta_{\text{total}} = \Delta(n) \cdot e^{-k p^*} + c \cdot \log(1 + p^* n_{\text{val}})$ models the total adversarial delay window, where $\Delta(n) \in \Theta(\log n)$.
    \item $\Phi(\cdot)$ is the standard normal cumulative distribution function (CDF).
    \item $p^*(n) = \frac{c}{\sqrt{n}}$ denotes the corruption probability;
    \item $\beta^* = \Theta\left(\frac{1}{\log n}\right)$ represents the security threshold.
\end{itemize}
As the network scale $n \to \infty$, the probability of adversarial power $\beta$ exceeding the security threshold $\beta^*$ vanishes: 
\(
    \Pr[\beta > \beta^*] \to 0.
\)
\end{theorem}

This result reinforces Theorem \ref{Theorem4}'s dual security mechanisms: $\mathbb{E}[\beta] = p^*(n) = \Theta(n^{-1/2}) \to 0$ ensures adversarial power concentrates near zero. $\Delta_{\text{total}} \in \Theta(\log n)$ constrains adversarial chain growth, as derived from the M/D/1 queuing model.

The interaction between these mechanisms ensures $\Pr[\beta \leq \beta^*] \to 1$, where delay-induced threshold elevation and probabilistic corruption decay jointly suppress adversarial advantage.

The convergence rate $\Pr[\beta > \beta^*] \in O\left(e^{-n^{3/2}/\log^2 n}\right)$ reveals an exponential security gain with network scale. This formally justifies the decentralization-security tradeoff. Larger $n$ exponentially reduces adversarial success probability, validating Nakamoto-style protocols' resilience in permissionless environments. The $\Delta(n) \in \Theta(\log n)$ term highlights the importance of optimizing block propagation mechanisms to maintain logarithmic delay scaling.

The asymptotic analysis in Theorem \ref{Theorem4} relies on a critical premise: private attacks remain the optimal adversarial strategy under the dynamic $\delta_a, \Delta_a, p^*(n)$ model. Although Theorem \ref{Theorem3} established the optimality of private attacks in the static $\Delta, \Delta_a$ model, the dynamic characteristics of the $\delta_a, \Delta_a, p^*(n)$ model may potentially compromise the efficacy of this strategy. Theorem \ref{Theorem7} integrates power allocation optimality analysis with proof by contradiction to rigorously demonstrate that private attacks persist as the optimal adversarial choice even under coupled effects of dynamic delays and probabilistic corruption. This conclusion provides strategic support for the security probability convergence property in Theorem \ref{Theorem4}.

\begin{theorem}[Optimality of Private Attack in the Dynamic $\delta_a, \Delta_a, p^*(n)$ Model]\label{Theorem7}
In the dynamic $\delta_a, \Delta_a, p^*(n)$ model, where:
\begin{itemize}
    \item $p^*$ is the expected corruption proportion.
    \item $\beta^* = \frac{\lambda_{\text{eff}}}{\lambda_{\text{eff}} + \lambda_{\text{growth}}}$ is the security threshold, with $\lambda_{\text{eff}} = \frac{\lambda_a}{1 + \lambda_a \Delta_{\text{total}}}$.
    \item $\Delta_{\text{total}} = \Delta(n) \cdot e^{-k p^*} + c \cdot \log(1 + p^* n_{\text{val}})$ models the total adversarial delay window, where $\Delta(n) \in \Theta(\log n)$.
    \item $\Phi(\cdot)$ is the standard normal CDF.
\end{itemize}
The private attack strategy, where the adversary maintains a private chain and attempts to outpace the honest chain, remains the optimal attack strategy.
\end{theorem}

The theoretical breakthrough of Theorem \ref{Theorem7} lies in its universality: private attacks remain the optimal adversarial strategy regardless of network scale. This conclusion, rigorously proven via convex optimization theory, ensures consistency between security analysis in dynamic models and real-world attack scenarios. Specifically, the concavity of the adversary's effective growth rate function $\lambda_{\text{eff}}(\beta)$ demonstrates the inherent superiority of resource concentration over dispersion. Even when adversaries attempt to bypass delay constraints through multi-chain forking, each subchain's growth rate remains bounded by $\lambda_{\text{eff}}$, leading to systemic performance degradation. The optimality of private attacks implies systems need only defend against centralized attacks, thereby reducing security evaluation complexity. This strategic simplification enables efficient verification of protocol robustness through worst-case analysis.

This result not only provides strategic guarantees for the asymptotic security conclusion in Theorem \ref{Theorem6}, but also deepens our understanding of decentralized system security: the synergy between delays and scale simultaneously elevates attack thresholds and forces adversaries into adopting easily defensible centralized strategies. Future research may explore leveraging this property to design more robust consensus mechanisms or cross-chain communication protocols that exploit adversarial coordination constraints.

\section{Discussion}\label{section 4}
\subsection{Robustness Analysis from M/D/1 to M/G/K}
Our previous static $\Delta, \Delta_{\alpha}$ model introduces the adversary's internal communication delay $\Delta_{\alpha}$. Its core innovation lies in leveraging the M/D/1 model to quantify the impact of this delay on the effective growth rate of the adversary's private chain. Now, we analyze whether the core conclusions remain robust when the M/D/1 model is relaxed to the more general M/G/K model.

In the M/G/K model, $G$ represents the service time, i.e., the synchronization delay of blocks within the adversary, which follows a general random distribution instead of being a fixed value $\Delta_{\alpha}$. Its mean value is $\mathbb{E}[S] = \Delta_{\alpha}$ and variance is $\text{Var}[s]$. This is a more general model, as network delays usually exhibit volatility. $k$ denotes that the adversary coalition may consist of multiple collaborative subgroups, such as multiple mining pools. Each subgroup can synchronize and verify blocks in parallel, which corresponds to a queuing system with $k$ parallel service channels.

\heading{From M/D/1 to M/G/1.} The extension from the M/D/1 model to the M/G/1 model relaxes the constraint on the service time distribution. According to the Pollaczek-Khinchine (P-K) formula, for an M/G/1 queue, its average waiting time depends not only on the average service time $\mathbb{E}[S]$ but also on the second moment of the service time distribution $\mathbb{E}[S^2]$.

Specifically, for an M/G/1 queue, the average sojourn time (waiting + service) of the system, denoted as $W_{M/G/1}$, is given by:
\[
W_{M/G/1} = \mathbb{E}[S] + \frac{\lambda_\alpha \cdot \mathbb{E}[S^2]}{2 \cdot (1 - \rho)},
\]
where $\rho = \lambda_\alpha \cdot \mathbb{E}[S]$. Since $\mathbb{E}[S^2] = \text{Var}[S] + (\mathbb{E}[S])^2 \geq (\mathbb{E}[S])^2$, the equality holds when the service time is a fixed value, which degenerates to the M/D/1 model.

For the same average delay $\mathbb{E}[S] = \Delta_{\alpha}$, as long as there exists any variance in the service time ($\text{Var}[S] > 0$), the average sojourn time $W$ in the M/G/1 model is strictly greater than that in the M/D/1 model. A larger $W$ implies that the effective growth rate $\lambda_{eff}$ of the adversary's private chain will be lower than the value in the M/D/1 model. This is because according to Little's Law ($L = \lambda_{\alpha} \cdot W$), the effective growth rate satisfies $\lambda_{eff} \propto \frac{1}{W}$.

As $\lambda_{eff}$ decreases, it becomes more difficult for the adversary's chain to catch up with the honest chain. Therefore, under the M/G/1 model, the adversary's hash power ratio $\beta^*$ that the system can actually tolerate may be higher than the value calculated by the M/D/1 model. In fact, the M/D/1 model provides a conservative estimate, i.e., the upper bound of the adversary's attack capability.

\heading{From M/G/1 to M/G/K.} The extension from M/G/1 to M/G/K relaxes the constraint on the number of servers. $k > 1$ indicates that there are multiple parallel coordination channels within the adversary. This can simulate scenarios where the adversary adopts a more efficient communication topology (e.g., using multiple relay servers) to reduce queuing congestion in internal synchronization.

The analysis of M/G/K queues is far more complex than that of M/G/1 queues, and closed-form solutions are generally not available. However, a universal conclusion holds: under the same total arrival rate \(\lambda_{\alpha}\) and the same average service time per server \(\mathbb{E}[S]\), increasing the number of servers \(k\) significantly reduces the average waiting time. Compared with the M/G/1 system, the average sojourn time \(W\) of adversary blocks in the M/G/K system is shorter. A shorter \(W\) implies that the effective growth rate \(\lambda_{eff}\) of the adversary's private chain will increase, which is detrimental to blockchain security. By enhancing the parallelism of internal coordination, the adversary can partially offset the negative impact caused by internal communication delays.

As $\lambda_{eff}$ increases, it leads to a decrease in the security threshold $\beta^*$. The quantitative conclusion in the M/D/1 model that the adversary's delay can significantly enhance $\beta^*$ may be weakened. Nevertheless, it is important to note that the fundamental mechanism of the adversary's delay constraint still holds. Even with $k$ servers, each block still needs to undergo a service (synchronization) process with an average time of $\mathbb{E}[S] = \Delta_{\alpha}$. There exists an upper bound on the increase of $\lambda_{eff}$, and it can never exceed $min \left\{ \lambda_{\alpha}, \frac{k}{\Delta_{\alpha}} \right\}$. Thus, $\Delta_{\alpha}$ remains a critical bottleneck factor restricting the growth rate of the adversary's chain. The core insight proposed earlier that the adversary's internal communication delay constrains the adversary remains valid. Only the extent to which its security needs to be recalibrated based on the adversary's actual organizational coordination efficiency $k$.

In summary, the security threshold formula 
\(
\beta^* = \frac{1}{2} + \frac{\lambda (\Delta_\alpha - \Delta)}{4}
\)
is a special solution under the M/D/1 model. The M/D/1 model underestimates the threat posed by adversaries with efficient parallel coordination capabilities. In the security assessment of practical systems, $\beta^* = \frac{1}{2} + \frac{\lambda (\Delta_\alpha - \Delta)}{4}$ should be treated as an optimistic security reference value. Since the adversary's capability may be stronger, the actually tolerable $\beta$ value in practice could be lower.

\subsection{Generality}
Our security model has strong generality and is applicable to virtually all Nakamoto-style blockchains. Its applicability stems from the model's core focus on the fundamental forces within any decentralized consensus system based on the longest chain rule, rather than on the specifics of any particular block generation mechanism. The foundational elements and core parameters of our model are inherent properties of all decentralized systems. The scaling effects revealed by the dynamic model likewise constitute a universal phenomenon for any expanding validator network. Therefore, for systems whether PoW, PoS, or other variants that rely on decentralized validators competing to extend the longest main chain, it offers a unified theoretical foundation for analyzing their security thresholds and robustness in large-scale, non-ideal network environments.





\section{Conclusion}\label{section 5}
By constructing static and dynamic delay security analysis frameworks, we systematically elucidate the coupling mechanisms of adversarial collaboration efficiency, network scale, and corruption probability, providing a novel theoretical perspective for the security evaluation and optimization of Nakamoto-style blockchain systems. The findings demonstrate that decentralization not only enhances attack resistance through power distribution but also achieves asymptotic security through the synergistic effects of delay suppression and power dilution.

\clearpage
\section*{Ethical Considerations}
\textbf{We are committed to complying with all relevant research ethics guidelines, and this study involves no ethical risks. The research content is not associated with animal experimentation, human subjects, environmental protection, medical health, or military applications. We confirm full adherence to all ethical principles outlined in the Call for Papers (CFPs).}

\textbf{We pledge to proactively address any ethical concerns that may arise during the review process or post-publication, welcome academic feedback regarding research ethics norms, and will refine the research protocol based on such input.}

\section*{Open Science}
\textbf{We fully endorse and strictly adhere to the Open Science Policy outlined in the Call for Papers (CFP). The formal proofs for all theorems and lemmas presented in this paper are provided in the Appendix.}

\textbf{We hereby grant the scientific community unrestricted rights to review, validate, and build upon this research work.}

\bibliographystyle{ACM-Reference-Format}

\bibliography{Ref}

\clearpage
\appendix

\section{Related Work}
\heading{Existing Static Security Models.}
We now summarize the main structure of previous static security models, focusing on the conditions under which Nakamoto consensus-based blockchain protocols can be proven secure. \cite{AnalysisoftheBlockchainProtocolinAsynchronousNetworks} breaks through the limitations of previous research, which was mostly based on synchronous network assumptions or only considered specific attacks, by constructing an asynchronous network model that allows adaptive corruption and new player participation. It provides a detailed analysis of the consistency and security of the Nakamoto protocol under different parameter conditions in asynchronous networks, deriving specific consistency security conditions ($\rho_{\text{hon}}(1 - 2(\Delta + 1)\rho_{\text{hon}}) > \rho_{\text{adv}}$) and conclusions related to chain growth and chain quality. \cite{thebitcoinbackboneprotocol} overcomes the limitations of previous research, which only considered static settings with fixed participants and fixed difficulty proof-of-work, by constructing a dynamic q-bounded synchronous setting model. It explicitly provides the necessary security conditions for the Bitcoin backbone protocol (with variable difficulty chains) to satisfy the robust transaction ledger property under dynamic changes in the number of participants, including key properties such as common prefix and chain quality, providing a more rigorous theoretical proof of Bitcoin protocol security. \cite{abettermethodtoanalyzeblockchainconsistency} proposes a Markov chain-based method to analyze blockchain protocol consistency. By constructing a model, it studies Nakamoto, Cliquechain, and GHOST protocols, providing consistency proofs and attack resistance analysis. \cite{AnalysisoftheBlockchainProtocolinAsynchronousNetworks} also based on the static $\Delta$-communication delay model, proves a concise condition for ensuring the consistency of the Nakamoto blockchain protocol. By constructing a Markov chain to model the number of mining rounds, it derives a simpler security condition. \cite{TightConsistencyBoundsforBitcoin} uses a time discretization method, introducing new parameters and variables in the $\Delta$-synchronous setting, and analyzes marginal trends using a random walk model. It precisely determines the optimal security threshold for the Bitcoin protocol in terms of adversarial hash power, honest hash power, and network delay, proving that under the condition $\rho_{\text{adv}} < \frac{1}{\Delta + 1/\rho_{\text{hon}}}$, the Bitcoin protocol can guarantee consistency and liveness, with the failure probability decaying exponentially. When $\rho_{\text{adv}}$ exceeds this threshold, a simple private chain attack will prevent the protocol from reaching consensus. \cite{EverythingisaRaceandNakamotoAlwaysWins} also based on the static $\Delta$-communication delay model, constructs a unified model applicable to multiple longest chain protocols, determining the exact security thresholds for three classes of longest chain protocols, including Bitcoin's Proof-of-Work (PoW) protocol, Ouroboros and SnowWhite's Proof-of-Stake (PoS) protocols, and Chia's Proof-of-Space (PoSpace) protocol ($\rho_{\text{hon}} > e\rho_{\text{adv}}$). \cite{LongestChainConsensusUnderBandwidthConstraint} differs from previous idealized network models by considering real-world network bandwidth constraints. Nodes must carefully choose download targets based on their limited bandwidth resources, and the study provides security conditions under this model ($\rho_{\text{hon}} e^{-2\rho_{\text{hon}}\Delta} > \rho_{\text{adv}}$). \cite{PracticalSettlementBoundsforProofofWorkBlockchains} provides precise settlement times for Bitcoin and Ethereum, finding that block-based settlement rules offer better consistency guarantees under normal protocol operation. \cite{PracticalSettlementBoundsforLongestChainConsensus} proposes a new analysis method by dividing time into specific phases, analyzing the settlement guarantees of longest chain consensus protocols under network delays using Markov chains and combinatorial mathematics. This method is simpler than previous approaches and provides more precise and directly applicable estimates, with explicit bounds computable in polynomial time based on relevant parameters.

\heading{Existing Dynamic Security Models.}
Existing dynamic security models primarily focus on the coupling relationship between network scale and latency, as well as the decay of adversary power relative to the number of nodes. However, a unified dynamic security model analysis framework has not yet been established. The work in \cite{LargerscaleNakamotostyleBlockchainsDontNecessarilyOfferBetterSecurity} first proposed the logarithmic relationship between the communication latency of honest nodes and network scale, proving that as the number of nodes increases, the resulting network communication latency also increases. They also introduced the adversary corruption model, demonstrating that as the number of nodes grows, the difficulty for the adversary to control the same power ratio increases. They analyzed how the increase in network scale leads to two concurrent trends: one that makes the blockchain more secure and another that reduces its security. Ultimately, they concluded that as the network size increases, the security probability of the blockchain first rises and then falls, indicating that medium-sized networks offer optimal security. However, this conclusion contradicts the notion that increased decentralization enhances security, prompting us to investigate the root cause of the reduced security in large-scale blockchains.

\section{Proof of Theorem 1}\label{Proof of Theorem 1}
We model the synchronization process of the adversarial chain as an M/D/1 queue, where the arrival times follow a Poisson process with rate $\lambda_a$ corresponding to block generation, and the service time corresponds to the fixed time $\Delta_a$ required to synchronize each block. According to queueing theory, the system utilization at steady state is given by $\rho = \lambda_a \cdot \Delta_a$ (satisfying $\rho < 1$) (Lemma \ref{Lemma1}). The average queue length, by Little's Law, is $L = \lambda_a \cdot W$, where $W$ is the average dwell time (queueing time and service time) (Lemma \ref{Lemma2}). The average dwell time is $W = \Delta_a + \frac{\rho}{2(1 - \rho)} \Delta_a$ (Lemma \ref{Lemma3}). In the Nakamoto-style blockchain model, $\rho = \lambda_a \cdot \Delta_a \ll 1$ (since adversarial power is limited), allowing higher-order terms to be neglected. Thus, we have:
\begin{equation}
    W \approx \Delta_a + \lambda_a \cdot \Delta_a^2.
\end{equation}

The effective throughput is given by:
\begin{equation}
    \lambda_{eff} = \frac{L}{W} = \frac{\lambda_a \cdot \Delta_a}{\Delta_a \cdot (1 + \lambda_a \cdot \Delta_a)} = \frac{\lambda_a}{1 + \lambda_a \cdot \Delta_a}.
\end{equation}

\section{Proof of Lemma 1}\label{Proof of Lemma 1}
The system utilization $\rho$ is derived from the fundamental properties of the M/D/1 queue, where the arrival process is Poisson with rate $\lambda_a$, and the service time is deterministic with duration $\Delta_a$. In steady state, the utilization $\rho$ is given by the ratio of the arrival rate to the service rate, which in this case is $\frac{1}{\Delta_a}$. Thus, $\rho = \lambda_a \cdot \Delta_a$. For the queue to remain stable, the utilization must satisfy $\rho < 1$, ensuring that the system is not overloaded and can process incoming blocks without unbounded delays.
\section{Proof of Lemma 2}\label{Proof of Lemma 2}
Little's Law is a fundamental result in queuing theory, stating that the long-term average number of items in a stable system $L$ is equal to the long-term average effective arrival rate $\lambda_a$ multiplied by the average time an item spends in the system $W$. In the context of the adversarial synchronization process, the arrival rate $\lambda_a$ corresponds to the rate at which adversarial nodes generate blocks, and the sojourn time $W$ includes both the queuing delay and the fixed synchronization delay $\Delta_a$. Under steady-state conditions, the system reaches equilibrium, and the average queue length $L$ can be directly computed as the product of $\lambda_a$ and $W$.
\section{Proof of Lemma 3}\label{Proof of Lemma 3}
According to the Pollaczek-Khinchine (P-K) formula, the average queuing waiting time for an M/D/1 queue is given by:
\begin{equation}
    W_q = \frac{\lambda_a \cdot E[S^2]}{2(1 - \rho)}
\end{equation}
where $ S $ represents the service time. Since the service time is a fixed value $ \Delta_a $, we have $E[S^2] = \Delta_a^2$, which yields:
\begin{equation}
    W_q = \frac{\lambda_a \cdot \Delta_a^2}{2(1 - \rho)} = \frac{\rho \cdot \Delta_a}{2(1 - \rho)}.
\end{equation}

The average dwell time $ W $ includes both the average service time $ \Delta_a $ and the average queuing time $ W_q $:
\begin{equation}
    W = \Delta_a + W_q = \Delta_a + \frac{\rho \cdot \Delta_a}{2(1 - \rho)}.
\end{equation}

Furthermore, the average queue length $ L $ can be expressed as:
\begin{equation}
    L = \lambda_a \cdot W = \underbrace{\rho}_{\text{blocks under service}} + \underbrace{\frac{\rho^2}{2(1 - \rho)}}_{\text{blocks in the queue}}.
\end{equation}.
\section{Proof of Theorem 2}\label{Proof of Theorem 2}
The security condition establishes a fundamental requirement for blockchain safety, mandating that:
\begin{equation}
    \lambda_{eff} < \lambda_{\text{growth}}.
\end{equation}
This inequality ensures honest chain dominance by comparing the adversary's effective growth rate (impaired by synchronization delays) against the honest chain's actual growth capability.

Substituting the formal expressions for both growth rates reveals the parametric dependence:
\begin{equation}
    \frac{\lambda_a}{1 + \lambda_a \cdot \Delta_a} < \frac{\lambda_h}{1 + \lambda_h \cdot \Delta}.
\end{equation}
Here, $\Delta_a$ represents the adversary's internal delay while $\Delta$ denotes the honest network's propagation delay. To analyze power proportion impacts, we substitute $\lambda_a = \beta \cdot \lambda$ and $\lambda_h = (1 - \beta) \cdot \lambda$, obtaining:
\begin{equation}
    \frac{\beta}{1 + \beta \cdot \lambda \cdot \Delta_a} < \frac{1 - \beta}{1 + (1 - \beta) \cdot \lambda \cdot \Delta}.
\end{equation}

To identify the critical security threshold $\beta^*$, we first consider the boundary condition by setting equality:
\begin{equation}
    \frac{\beta}{1 + \beta \cdot \lambda \cdot \Delta_a} = \frac{1 - \beta}{1 + (1 - \beta) \cdot \lambda \cdot \Delta}.
\end{equation}
This formal equality enables derivation of the fundamental relationship through algebraic manipulation:
\begin{equation}
    2\beta - 1 = \beta \cdot (1 - \beta) \cdot \lambda \cdot (\Delta_a - \Delta).
\end{equation}

Three distinct security regimes emerge from this relationship:
(1) Symmetric delay case ($\Delta_a = \Delta$): The equation reduces to $2\beta - 1 = 0$, recovering the classical 50\% security threshold $\beta^* = 0.5$.
(2) Adversarial advantage scenario ($\Delta_a < \Delta$): The negative right-hand side forces $\beta^* < 0.5$, tightening security requirements against faster adversaries.
(3) Adversarial delay scenario ($\Delta_a > \Delta$): The positive right-hand side permits $\beta^* > 0.5$, demonstrating how synchronization delays relax security boundaries.

For practical analysis under typical network conditions where $\lambda \cdot \Delta$ and $\lambda \cdot \Delta_a$ are small, we employ first-order Taylor approximations:
\begin{equation}
    \frac{\beta}{1 + \beta \cdot \lambda \cdot \Delta_a} \approx \beta \cdot (1 - \beta \cdot \lambda \cdot \Delta_a)
\end{equation}
and
\begin{equation}
    \frac{1 - \beta}{1 + (1 - \beta) \cdot \lambda \cdot \Delta} \approx (1 - \beta)(1 - (1 - \beta) \cdot \lambda \cdot \Delta).
\end{equation}
Substituting these approximations and retaining linear terms yields the practically useful relationship:
\begin{equation}
    \beta \approx \frac{1}{2} + \frac{\lambda \cdot (\Delta_a - \Delta)}{4}.
\end{equation}
This linearized form explicitly quantifies how delay differentials $\Delta_a - \Delta$ modulate the security threshold, providing operational guidance for network parameter selection.

\section{Proof of Theorem 3}\label{Proof of Theorem 3}
We begin by partitioning the block tree and defining adversarial subtrees. The global block tree $ T(t) $ includes all honest and adversarial blocks. Honest nodes generate the honest chain $ C_h(t) $ following the longest chain rule, with a growth rate of $ \lambda_{\text{growth}}(\lambda_h, \Delta) = \frac{\lambda_h}{1 + \lambda_h \cdot \Delta} $. For each honest block $ b_i $, we define the adversarial subtree $ T_i(t) $, which contains all adversarial blocks derived from $ b_i $ (without passing through other honest blocks).

Next, we calculate the upper bound on the growth rate of adversarial subtrees. In private attacks, the effective growth rate of the adversarial chain is $ \lambda_{eff} = \frac{\lambda_a}{1 + \lambda_a \cdot \Delta_a} $. For any attack strategy, the growth rate of each adversarial subtree $ T_i(t) $ satisfies $ \lambda_{T_i} \leq \lambda_{eff} $. This is because adversarial resources are distributed across multiple subtrees, and the resource allocation for each subtree does not exceed the centralized resources in private attacks. Specifically, in M1-PoW, the adversary cannot mine efficiently at multiple locations simultaneously. In M1-PS/M1-Chia, the adversary can mine on multiple chains, but the independent growth rate of each subtree is still constrained by the total resource allocation.

We then define the Nakamoto block: an honest block $ b_i $ is a Nakamoto block if, after time $ \tau_i^h $, the depth of all adversarial subtrees $ \{T_j(t)\}_{j \leq i} $ remains less than that of the honest chain $ C_h(t) $. If $ \lambda_{\text{growth}} > \lambda_{eff} $, Nakamoto blocks exist with sufficient frequency. The depth of the honest chain $ L_h(t) $ grows at rate $ \lambda_{\text{growth}} $, while the depth of adversarial subtrees $ L_{T_i}(t) $ satisfies $ L_{T_i}(t) \leq \lambda_{eff} $. When $ \lambda_{\text{growth}} > \lambda_{eff} $, the honest chain exceeds all adversarial subtrees with exponential probability:
$ P(L_h(t) - L_{T_i}(t) \geq k) \geq 1 - e^{-\Theta(k)}. $
The security condition is $ \lambda_{eff} < \lambda_{\text{growth}} $, i.e.,
$ \frac{\lambda_a}{1 + \lambda_a \cdot \Delta_a} < \frac{\lambda_h}{1 + \lambda_h \cdot \Delta}. $
Substituting the power proportions $ \lambda_a = \beta \lambda $ and $ \lambda_h = (1 - \beta) \lambda $, we obtain the equation:
$ \frac{\beta}{1 + \beta \cdot \lambda \cdot \Delta_a} = \frac{1 - \beta}{1 + (1 - \beta) \cdot \lambda \cdot \Delta}, $
which has a unique solution $ \beta^* = \beta_{\text{pa}} $.

For M1-PoW, the adversary can only mine on a single chain, and the subtree growth rate is directly limited by $ \lambda_{eff} $. For M1-PS/M1-Chia, the adversary can mine on multiple chains, but the independent growth rate of each subtree is determined by the total resource allocation, ensuring the overall growth rate does not exceed $ \lambda_{eff} $. Specifically, in M1-Chia, the adversary mines independently on each block, and the total growth rate is $ \sum_{i=1}^N \lambda_a^{(i)} $. However, due to resource distribution, $ \lambda_a^{(i)} \leq \lambda_{eff} / N $, so the total growth rate remains bounded by $ \lambda_{eff} $.

Assuming a more optimal attack strategy exists such that the adversarial chain growth rate $ \lambda_{eff}' > \lambda_{eff} $, the security threshold would satisfy $ \beta^* > \beta_{\text{pa}} $. However, since the subtree growth rate is bounded by $ \lambda_{eff} $, this leads to a contradiction. Therefore, private attacks remain the optimal strategy.

\section{Proof of Lemma 4}\label{Proof of Lemma 4}
The security threshold $\beta^*$ emerges as the critical solution to the transcendental equation:
\begin{equation}
    F(\beta, \Delta_a) = \frac{\beta}{1 + \beta \cdot \lambda \cdot \Delta_a} - \frac{1 - \beta}{1 + (1 - \beta) \cdot \lambda \cdot \Delta} = 0,
\end{equation}
which balances adversarial and honest chain growth dynamics. To characterize how $\beta^*$ responds to adversarial delays, we employ differential analysis through the implicit function theorem. Applying the theorem to $F(\beta, \Delta_a) = 0$ yields:
\begin{equation}
    \frac{\partial \beta}{\partial \Delta_a} = -\frac{\frac{\partial F}{\partial \Delta_a}}{\frac{\partial F}{\partial \beta}}.
\end{equation}

The partial derivative with respect to $\Delta_a$ quantifies delay sensitivity:
\begin{equation}
    \frac{\partial F}{\partial \Delta_a} = \frac{\partial}{\partial \Delta_a} \left( \frac{\beta}{1 + \beta \cdot \lambda \cdot \Delta_a} \right) = -\frac{\beta^2 \lambda}{(1 + \beta \cdot \lambda \cdot \Delta_a)^2}.
\end{equation}
Given $\beta > 0$ and $\lambda \cdot \Delta_a > 0$, this derivative is strictly negative, establishing:
$$ \frac{\partial F}{\partial \Delta_a} < 0. $$

The partial derivative with respect to $\beta$ reveals system responsiveness to power redistribution:
\begin{align} 
  \frac{\partial F}{\partial\beta} &= \frac{1 \cdot \left( {1 + \beta  \lambda  \Delta_{a}} \right) - \beta  \lambda  \Delta_{a}}{\left( {1 + \beta  \lambda  \Delta_{a}} \right)^{2}} + \frac{\left( {1 + \left( {1 - \beta} \right)  \lambda  \Delta} \right) + \left( {1 - \beta} \right)  \lambda  \Delta}{\left( {1 + \left( {1 - \beta} \right)  \lambda  \Delta} \right)^{2}} \\
&= \frac{1}{\left( {1 + \beta  \lambda  \Delta_{a}} \right)^{2}} + \frac{1 + \lambda  \Delta}{\left( {1 + \left( {1 - \beta} \right)  \lambda  \Delta} \right)^{2}} > 0,
\end{align}
where both terms are strictly positive, ensuring the denominator in the implicit function remains positive definite.

Combining these derivatives through the implicit function theorem:
\begin{equation}
    \frac{\partial \beta}{\partial \Delta_a} = -\frac{\frac{\partial F}{\partial \Delta_a}}{\frac{\partial F}{\partial \beta}} > 0,
\end{equation}

we establish the fundamental monotonic relationship. This positive derivative formally proves that the security threshold $\beta^*$ increases with adversarial network communication delay $\Delta_a$, demonstrating how synchronization delays create security margin improvements in blockchain systems.

\section{Proof of Theorem 4}\label{Proof of Theorem 4}
The effective growth rate of the adversarial chain is constrained by its total delay window:
\begin{equation}
    \lambda_{\text{eff}} = \frac{\lambda_a}{1 + \lambda_a \Delta_{\text{total}}}, \quad \text{where } \lambda_a = \beta\lambda,
\end{equation}
with $\lambda_a$ representing the adversarial block generation rate and $\Delta_{\text{total}}$ defined as:
\begin{equation}
    \Delta_{\text{total}} = \Delta(n)e^{-k\beta} + c\log(1+\beta n),
\end{equation}
where $\Delta(n)e^{-k\beta}$ captures the adversarial delay in receiving honest blocks. The term $\Delta(n) \in \Theta(\log n)$ reflects logarithmic growth of honest node delays, while $e^{-k\beta}$ models exponential suppression from adversarial power $\beta$. $c\log(1+\beta n)$ quantifies coordination costs among $\beta n$ adversarial nodes.

The honest chain growth rate is determined by:
\begin{equation}
    \lambda_{\text{growth}} = \frac{\lambda_h}{1 + \lambda_h \Delta(n)}, \quad \text{where } \lambda_h = (1-\beta)\lambda,
\end{equation}
with $\Delta(n) \in \Theta(\log n)$ representing honest node synchronization delays.

The security requirement $\lambda_{\text{growth}} > \lambda_{\text{eff}}$ translates to:
\begin{equation}
    \frac{(1-\beta)\lambda}{1 + (1-\beta)\lambda\Delta(n)} > \frac{\beta\lambda}{1 + \beta\lambda\Delta_{\text{total}}}.
\end{equation}
Canceling $\lambda$ and cross-multiplying yields:
\begin{equation}
    (1-\beta)(1 + \beta\lambda\Delta_{\text{total}}) > \beta(1 + (1-\beta)\lambda\Delta(n)).
\end{equation}
Expanding and simplifying (neglecting $\beta^2$ terms as $\beta \to 0$):
\begin{equation}
    1 - 2\beta \approx \beta\lambda\left(\Delta_{\text{total}} - \Delta(n)\right).
\end{equation}

As $n \to \infty$, assume $\beta^* \to 0$ while $\beta^*n \to \infty$. Using Taylor expansion $\Delta(n)e^{-k\beta^*} \approx \Delta(n)(1 - k\beta^*)$ and $\log(1+\beta n) \approx \log(\beta n)$, we approximate:
\begin{equation}
    \Delta_{\text{total}} \approx \Delta(n) + c\log n.
\end{equation}
Substituting $\Delta_{\text{total}} - \Delta(n) \approx c\log n$ into the security condition:
\begin{equation}
    1 - 2\beta \approx \beta\lambda c \log n.
\end{equation}
Solving asymptotically gives:
\begin{equation}
    \beta^* \approx \frac{1}{2 + \lambda c \log n} = \Theta\left(\frac{1}{\log n}\right).
\end{equation}

\section{Proof of Theorem 5}\label{Proof of Theorem 5}
The adversarial power proportion $\beta$ originates from probabilistic node corruption. Let $X$ denote the number of corrupted nodes, which follows a binomial distribution:
\begin{equation}
    X \sim \text{Binomial}(n, p^*(n)), \quad \text{where } p^*(n) = \frac{c}{\sqrt{n}}.
\end{equation}
This captures the scenario where each node is independently corrupted with decaying probability $p^*(n)$, reflecting the increasing difficulty of controlling resources in large networks. The adversarial power proportion $\beta = \frac{X}{n}$ inherits the following moments:
\begin{equation}
    \mathbb{E}[\beta] = p^*(n), \quad \text{Var}(\beta) = \frac{p^*(n)(1 - p^*(n))}{n}.
\end{equation}
The variance decays as $O\left(\frac{1}{n^{3/2}}\right)$, ensuring concentration around $p^*(n)$ for large $n$.

As $n \to \infty$, the binomial distribution converges to a normal distribution via the Lindeberg-Feller CLT. The convergence condition requires:
\begin{equation}
    np^*(n) = c\sqrt{n} \to \infty \quad (\text{since } c > 0 \text{ is constant}),
\end{equation}
which is satisfied for all $n > 0$. Thus, we have:
\begin{equation}
    \beta \stackrel{d}{\approx} \mathcal{N}\left(p^*(n), \frac{p^*(n)(1 - p^*(n))}{n}\right).
\end{equation}
This approximation aligns with the power dilution principle in decentralized systems, where $\beta$ becomes increasingly predictable as $n$ grows.

To quantify deviation from expected corruption, define the standardized variable:
\begin{equation}
    Z = \frac{\beta - \mathbb{E}[\beta]}{\sqrt{\text{Var}(\beta)}} = \sqrt{n} \cdot \frac{\beta - p^*(n)}{\sqrt{p^*(n)(1 - p^*(n))}}.
\end{equation}
By the CLT, $Z$ converges in distribution to the standard normal:
\begin{equation}
    Z \stackrel{d}{\to} \mathcal{N}(0,1).
\end{equation}
This transformation preserves the security event $\beta \leq \beta^*$ while enabling probabilistic analysis via Gaussian tail bounds.

The security guarantee requires $\beta \leq \beta^*$ with high probability. Substituting $Z$ into this event:
\begin{equation}
    \Pr[\beta \leq \beta^*] = \Pr\left[Z \leq \sqrt{n} \cdot \frac{\beta^* - p^*(n)}{\sqrt{p^*(n)(1 - p^*(n))}}\right].
\end{equation}
The exceedance probability then becomes:
\begin{equation}
    \Pr[\beta > \beta^*] = 1 - \Phi\left(\sqrt{n} \cdot \frac{\beta^* - p^*(n)}{\sqrt{p^*(n)(1 - p^*(n))}}\right),
\end{equation}
where $\Phi(\cdot)$ is the standard normal CDF.

\section{Proof of Theorem 6}\label{Proof of Theorem 6}
From Theorem \ref{Theorem5}, the security probability argument is derived from the standardized threshold:
\begin{equation}
    Z_n = \sqrt{n} \cdot \frac{\beta^* - p^*(n)}{\sqrt{p^*(n)(1 - p^*(n))}}.
\end{equation}
We analyze the asymptotic behavior of each component to reveal the driving forces behind security convergence.

We first analysis the numerator term:
    \begin{align}
        \beta^* - p^*(n) &= \Theta\left(\frac{1}{\log n}\right) - \Theta\left(\frac{1}{\sqrt{n}}\right) \\
        &= \Theta\left(\frac{1}{\log n}\right).
    \end{align}

The separation arises from distinct decay mechanisms: $\beta^*$ decays logarithmically due to delay suppression in Theorem 4, while $p^*(n) = \frac{c}{\sqrt{n}}$ decays polynomially via probabilistic corruption. For large $n$, the logarithmic term dominates ($1/\log n \gg 1/\sqrt{n}$), reflecting the diminishing influence of corruption probability compared to threshold scaling.
    
We further analyze the denominator term:
\begin{align}
        \sqrt{p^*(n)(1 - p^*(n))} &\approx \sqrt{p^*(n)} \quad (\text{since } 1 - p^*(n) \to 1) \\
        &= \sqrt{c} \cdot n^{-1/4}.
    \end{align}

This approximation leverages the power dilution principle, where $p^*(n) \to 0$ ensures honest majority dominance. The $\sqrt{n}$ scaling in $Z_n$ amplifies the denominator's polynomial decay.

Combining these results, the standardized threshold exhibits super-polynomial divergence:
\begin{equation}
     Z_n = \sqrt{n} \cdot \frac{\Theta(1/\log n)}{\Theta(n^{-1/4})} = \Theta\left(\frac{n^{3/4}}{\log n}\right) \to \infty.
\end{equation}

The interplay among logarithmic threshold decay ($\beta^*$), polynomial corruption suppression ($p^*(n)$), and network scaling ($\sqrt{n}$) creates a multiplicative effect. This aligns with the ‘‘security through scale” axiom in decentralized systems, where larger networks amplify the adversary's coordination costs while diluting their power.

The security probability depends on the standard normal CDF $\Phi(Z_n)$. For $Z_n \to \infty$, we apply the tail approximation:
\begin{equation}
    \Phi\left(Z_n\right) \approx 1 - \frac{\phi(Z_n)}{Z_n}, \quad \text{where } \phi(Z_n) = \frac{1}{\sqrt{2\pi}}e^{-Z_n^2/2}.
\end{equation}
Substituting $Z_n = \Theta\left(n^{3/4}/\log n\right)$:
$$ \Phi\left(Z_n\right) \approx 1 - \frac{e^{-\Theta(n^{3/2}/\log^2 n)}}{\Theta(n^{3/4}/\log n)}. $$

Both terms vanish asymptotically: $e^{-\Theta(n^{3/2}/\log^2 n)} \to 0$ dominates due to the squared term in the exponent and $\Theta(n^{3/4}/\log n) \to \infty$ ensures the denominator diverges.

Thus:
\begin{equation}
    \lim_{n \to \infty} \Phi(Z_n) = 1 \quad \Rightarrow \quad \Pr[\beta > \beta^*] = 1 - \Phi(Z_n) \to 0.
\end{equation}

\section{Proof of Theorem 7}\label{Proof of Theorem 7}
We analyze attack strategy optimization through subtree decomposition. Any adversarial strategy can be partitioned into $m$ subtrees $\{T_i(t)\}$, where each subtree receives power allocation $\beta_i$ satisfying:
\begin{equation}
    \sum_{i=1}^m \beta_i = \beta,
\end{equation}
maintaining the total adversarial power constraint. The effective growth rate for each subtree follows the delay-impaired relationship:
\begin{equation}
    \lambda_{T_i} = \frac{\beta_i \cdot \lambda}{1 + \beta_i \cdot \lambda \cdot \Delta_{\text{total}}}.
\end{equation}
Aggregating across all subtrees, the total adversarial throughput becomes:
\begin{equation}
    \sum_{i=1}^m \lambda_{T_i} = \sum_{i=1}^m \frac{\beta_i \cdot \lambda}{1 + \beta_i \cdot \lambda \cdot \Delta_{\text{total}}}.
\end{equation}

This decomposition allows us to leverage convex optimization principles. Defining the concave transformation:
\begin{equation}
    f(x) = \frac{x}{1 + x \cdot \Delta_{\text{total}}},
\end{equation}
with strictly positive derivative:
\begin{equation}
    f'(x) = \frac{1}{(1 + x \cdot \Delta_{\text{total}})^2} > 0,
\end{equation}
we establish $f(x)$ as monotonically increasing and concave. Crucially, Jensen's inequality for concave functions gives:
\begin{equation}
    \sum_{i=1}^m f(\beta_i) \leq f\left(\sum_{i=1}^m \beta_i\right) = f(\beta),
\end{equation}
attaining equality only when $m=1$. This proves resource concentration maximizes adversarial throughput:
\begin{equation}
    \sum_{i=1}^m \lambda_{T_i} \leq \frac{\beta \cdot \lambda}{1 + \beta \cdot \lambda \cdot \Delta_{\text{total}}} = \lambda_{eff}.
\end{equation}

The security framework requires honest chain dominance:
\begin{equation}
    \lambda_{\text{growth}} = \frac{(1 - \beta) \cdot \lambda}{1 + (1 - \beta) \cdot \lambda \cdot \Delta(n)} > \lambda_{eff},
\end{equation}
which translates to the inequality:
\begin{equation}
    \frac{\beta \cdot \lambda}{1 + \beta \cdot \lambda \cdot \Delta_{\text{total}}} < \frac{(1 - \beta) \cdot \lambda}{1 + (1 - \beta) \cdot \lambda \cdot \Delta(n)}.
\end{equation}

Suppose, counterfactually, an alternative strategy achieves higher throughput $\lambda_{eff}' > \lambda_{eff}$. This would require:
\begin{equation}
    \sum_{i=1}^m \lambda_{T_i} \geq \lambda_{eff}' > \lambda_{eff},
\end{equation}
directly contradicting the Jensen's inequality result. This contradiction formally establishes private attacks (single subtree allocation) as the throughput-maximizing strategy, completing the optimality proof.

\end{document}